\documentclass[twocolumn,showpacs,preprintnumbers,amsmath,amssymb,nofootinbib]{revtex4}
\usepackage{graphicx}
\usepackage{epsfig}
\newcommand{\beq}{\begin{eqnarray}}
\newcommand{\eeq}{\end{eqnarray}}
\begin{document}
\title{Theory of Magnetic Seed-Field
Generation during the Cosmological First-Order
Electroweak Phase Transition}
\author{Trevor Stevens\\
Department of Physics, West Virginia Wesleyan College, Buckhannon, West Virginia 26201\\
Mikkel B. Johnson \\
Los Alamos National Laboratory, Los Alamos, New Mexico 87545 \\}
\begin{abstract}
     We present a theory of the generation of magnetic seed fields in bubble collisions during a first-order electroweak phase transition (EWPT) possible for some choices of parameters in the minimal supersymmetric Standard Model.  The theory extends earlier work and is formulated to assess the importance of surface dynamics in such collisions. We are led to linearized equations of motion with $O(3)$ symmetry appropriate for examining collisions in which the Higgs field is relatively unperturbed from its mean value in the collision volume. Coherent evolution of the charged $W$ fields within the bubbles is the main source of the {\it em} current for generating the seed fields, with fermions also contributing through the conductivity terms. 
We present numerical simulations within this formulation to quantify the role of the surface of the colliding bubbles, particularly the thickness of the surface, and to show how conclusions drawn from earlier work are modified. The main sensitivity arises such that the steeper the bubble surface the more enhanced the seed fields become. Consequently, the magnetic seed fields may be several times larger and smoother over the collision volume than found in earlier studies. 
Our work thus provides additional support to the supposition that magnetic fields produced during the EWPT in the early universe seed the galactic and extra-galactic magnetic fields observed today. 

\end{abstract}

\maketitle
\noindent
\vspace{1mm}


\vspace{1mm}
\noindent

\section{Introduction}
Identifying the source of the observed large-scale galactic and extra-galactic
magnetic fields remains an unresolved problem of
astrophysics~\cite{gr}. One of the interesting possible sources is
cosmological magnetogenesis, where the seed fields would have arisen
during one of the early-universe phase transitions.  In this work our
interest is seed field production during the electroweak phase
transition (EWPT) during which the Higgs and the other particles acquired
their masses.

If magnetogenesis occurred during the EWPT, it most likely required a
first-order phase transition. A first-order phase transition proceeds by a process in which bubbles of matter in the broken phase nucleate within the unbroken phase, similar to the familiar process of steam condensing to water as the water-vapor mixture is cooled.  Although it is generally believed that
there can be no first-order EWPT in the Standard Model~\cite{klrs},
there has been a great deal of activity in supersymmetric
extensions~\cite{r90}, and for certain minimal extensions of the
Standard Model there can be a first-order phase
transition~\cite{laine,cline,losada}. Limits on parameter-space of
the minimal extension of the standard model (MSSM) placed by electric
dipole moment measurements and dark matter searches allow a
first-order EWPT which could lead to
successful electroweak baryogenesis~\cite{cpr} and the possibility
that we are exploring, namely that seed fields responsible
for the large-scale magnetic fields seen today are created during the
era of the EWPT.

Interest in these issues has led to quantitative studies of EWPT 
magnetogenesis based on the solution of equations of motion (EOM) 
derived from specific models.  In the Abelian Higgs model, the first-order phase transition developed as the Universe condensed into bubbles consisting of localized regions of space filled by Higgs field in a broken phase.  This model was one of the 
earliest attempts to describe seed field production during the EWPT.  The EOM related the 
seed fields to gradients in the phase of the Higgs field 
that were produced when bubbles merged 
following nucleation~\cite{kv95,ae98,cst00}. Simple and transparent 
solutions to the EOM evolved from specific field configurations applied at 
the point of collision in a relativistic $O(1,2)$ symmetric model.

The production mechanism of seed fields within an EOM 
approach has been pursued more recently within the framework of the MSSM~\cite{jkhhs,stevens1} along the lines of the 
Abelian Higgs model in Ref.~\cite{kv95,ae98,cst00} in which $O(1,2)$ symmetric solutions evolved from 
specific field configurations applied at 
the point of collision. In this work, new EOM were derived from the MSSM Lagrangian, and accordingly the bubbles developed a coherent mode of charged $W^\pm$ fields.  As the bubbles merged, the charged gauge fields replaced gradients in the phase of the Higgs field as the source of the  electromagnetic {\it em} currents producing the magnetic 
seed fields, and the mechanism by which this occurred was developed in detail. Numerical results~\cite{stevens1} showed that the
MSSM produced seed fields of a size similar to the Abelian Higgs model even though the source of the current in the two approaches was quite different.

Although these earlier studies gave insight into the production 
mechanism of seed fields in a first-order EWPT, applying boundary conditions at 
the time of the collision as implemented in the $O(1,2)$ formulations
did not permit an assessment of the role of the dynamics of the bubble 
surface in the seed field generation. To explore the role of the surface it is necessary to specify the values of the $W$ fields and their time derivatives on a surface $(t=t_0,r,z)$ {\it before} the collision occurs.
Thus, incorporating the initial stages of evolution of individual bubbles on the collision is an aspect of physics absent from the treatments found in Refs.~\cite{kv95,ae98,cst00,jkhhs,stevens1}. 

In our more recent studies~\cite{stevens3,stevens3a} results of calculations were presented in which bubble surface dynamics were taken into account. We identified a source of quantitative sensitivity to the bubble surface, and the results presented therein showed that the magnetic fields produced could be as large as, and possibly even larger than, those calculated in their absence. Encouraged by these results, in the present work we develop and extend the EOM formulation within the MSSM upon which this earlier work was based, along lines identified there.

We begin, in Sect.~II, by reviewing our EOM approach.  We also discuss
briefly the nature of a first-order EWPT and some of the issues associated with the dynamics of the bubble surface that our theory is intended to address. In Sect.~III we present our extended theory developed in $3+1$ dimensions in a regime where the bubble collisions may be considered "gentle"~\cite{jkhhs,stevens1}. For gentle collisions the EOM linearize and display a transparent connection to the earlier work of Refs.~\cite{kv95,ae98,cst00,stevens1}. 
Various theoretical considerations necessary for assessing 
the role of the bubble surface in magnetic field generation, including the importance of establishing appropriate initial conditions, are developed. 

Out theory is applied in a specific model along the lines of Ref.~\cite{stevens1,stevens3a} in Sect.~\ref{s:solution}. Because in our present formulation boundary conditions are applied  before the collision occurs, we are able to examine in addition to collisions, the nucleation process, which is the 
evolution of the bubbles before the collision takes place. The numerical solutions of nucleation and collisions in this model are then presented in the following two sections. 

In Sect.~\ref{s:thick} we examine the $W$ fields, the current produced in collisions, and the magnetic field. 
For collisions, find the magnetic field to be larger in both scale and magnitude  compared to our earlier $O(1,2)$ results~\cite{stevens1}. In Sect.~\ref{s:sens} we estimate the sensitivity of the seed fields to the steepness of the surface of the scalar field, and in Sect.~\ref{s:sensvsig} the sensitivity to the bubble wall speed and conductivity of the medium. Compared to results given in Refs.~\cite{stevens3a}, we find that the magnetic seed fields are not only larger in magnitude but extend over substantially larger spatial scales than the results shown there.

\section{ Magnetic field creation
during a first-order EWPT in the MSSM }
\label{ss:lagrangian}

The EOM of this work are based on the same underlying MSSM Lagrangian as that of Ref.~\cite{stevens1}.  This Lagrangian is assumed to support a 
first-order phase transition and is of the form
\beq
\label{L}
    {\cal L}^{MSSM} & = & {\cal L}^{1} + {\cal L}^{2} \nonumber  \\
        &+& {\rm leptonic,~quark,~and~supersymmetric} \nonumber \\
&&{ \rm partner~interactions } \nonumber \\
           {\cal L}^{1} & = & -\frac{1}{4}W^i_{\mu\nu}W^{i\mu\nu}
    -\frac{1}{4} B_{\mu\nu}B^{\mu\nu} \nonumber \\
   {\cal L}^{2} & = & |(i\partial_{\mu} -\frac{g}{2} \tau \cdot W_\mu
   - \frac{g'}{2}B_\mu)\Phi|^2 \nonumber \\
 &-&V(\Phi,T)~,
\eeq
where $T$ is the temperature and
\beq
\label{wmunu}
    W^i_{\mu\nu} & = & \partial_\mu W^i_\nu - \partial_\nu W^i_\mu
   - g \epsilon_{ijk} W^j_\mu W^k_\nu \nonumber \\
   B_{\mu\nu} & = & \partial_\mu B_\nu -  \partial_\nu B_\mu~.
\eeq
Here $W^i$, with i = (1,2), are the 
$W^+,W^-$ fields, $\Phi$ is
the Higgs field, and $\tau^i$ is the SU(2) generator. Fermions are
not explicitly considered since earlier work to which we want to compare likewise ignored them.  Because we are working within the framework of the MSSM, the bubbles that form as the phase transition progresses naturally consist of a region of space filed by the Higgs field along with a cloud of other constituents of the MSSM Lagrangian in the broken phase.

As in Ref.~\cite{stevens1}, we first derive``exact" EOM using an effective Lagrangian at the classical level from which the supersymmetric partners have been projected out as explicit degrees of freedom, but whose effect is retained by a  renormalization of the effective potential to maintain the properties of the first-order phase transition. These EOM are complicated nonlinear partial
differential equations coupling the $W$, $B$,
and $\Phi$ fields.
From their solution 
one may obtain the physical $Z$ and $A^{{\it em}}$ fields, 
\beq
\label{AZ}
     A^{em}_\mu &=& \frac{1}{\sqrt{g^2 +g^{'2}}}(g'W^3_\mu +g B_\mu) 
\nonumber \\
     Z_\mu &=& \frac{1}{\sqrt{g^2 +g^{'2}}}(g W^3_\mu -g' B_\mu) \; .
\eeq

In the picture we are developing, the Higgs field plays a central dynamical
role in EW bubble nucleation and collisions, and therefore the 
effective (now appropriately renormalized) Higgs potential $V(\Phi,T )$ is an essential element in the 
theory. Although this potential is not known at the present time,
depending as it does on the unknown parameters of the MSSM as well as
the properties of the plasma in the early Universe at the time of the
EWPT, its specific form is not relevant for the
purposes of this paper.  We require only that it should produce
a first-order phase transition, consistent with certain MSSM 
extensions including for example those with a light right-handed
Stop~\cite{bodeker}.  

The various parameters are discussed in many
publications~\cite{laine}. For our calculations we
use the laboratory values,
\beq
\label{cc}
g&=&e/\sin\theta_W = 0.646~, \nonumber \\
g'&=&g\tan\theta_W =0.343~, \nonumber \\
m_W&=&80.4~{\rm GeV}~,
\eeq
where $m_W$ is the mass of the $W^\pm$ bosons, and we define 
\beq
\label{capG}
G&\equiv& gg'/\sqrt{g^2 +g^{'2}} =0.303~.
\eeq
In this section and throughout the paper units are such that $\hbar=c=1$, with distance and time expressed in units of $m_W$.

\subsection{First-order electroweak phase transition}
\label{ss:firstew}

Coleman's model~\cite{coleman} provides a conceptual framework based on a Lagrangian for understanding the phase transition. Although oversimplified in that it lacks medium effects, which can lead to an asymptotic wall speed $v_{wall}<1$ depending on the pressure difference in the true and false vacuum, it is seminal in that it was one of the earliest EOM descriptions of the physics of a first order phase transition. In his model, prior to nucleation, the dynamics of bubbles is
formulated in the Euclidean metric, which is $O(4)$ symmetric.  
After nucleation, the bubble expands in $O(1,3)$-symmetric Minkowski 
space-time~\cite{coleman}.  As the phase transition develops the 
bubbles start to merge or "collide". Eventually they completely 
merge, at which point the phase transition is completed. In subsequent work magnetic fields have been understood to be 
created in the collisions of bubbles.

In the Lagrangian of Eq.~(\ref{L}), the EWPT is driven by nucleation of the scalar Higgs field just as imagined in Coleman's
model~\cite{coleman}.  In this picture, the vacuum state of the Universe
corresponds to a local
minimum in $V(\Phi ,T)$.  The phase transition occurs as
the temperature $T$ is lowered through the transition temperature
$T_c$ when $V(\Phi ,T)$ develops a degenerate second minimum at a
larger value of $<\Phi>$ separated from this minimum
by a barrier.  As the universe continues to expand and cool, the
depth of the second minimum increases, meaning that the Universe
can lower its energy by moving from the original, now metastable,
false vacuum to the lower energy true vacuum. Because the two minima
are separated by a barrier, the transition from the false to the true
vacuum is delayed as the temperature continues to drop, a process
referred to as supercooling.  This delay influences bubble characteristics, and a first-order phase transition is accordingly classified as weak or strong depending on the degree of supercooling. A
comprehensive phenomenological study of the kinetics of cosmological
first-order phase transitions, such as the EWPT, in
terms of such an effective potential is given for example in Ref.~\cite{eikr}.

\subsection{Bubble dynamics and magnetic field creation with $O(1,2)$ Symmetry}
\label{ss:bubandphase}

In their analysis of the Abelian Higgs model, Kibble and
Vilenkin~\cite{kv95} obtained magnetic fields as bubbles merge in a regime of
gentle collisions.  They obtained EOM in this case by making an expansion
about point $\rho(x)=\rho_0$ (the ``Kibble-Vilenkin point").  From these EOM, expressed in terms of
the variables $(z,\tau=\sqrt{t^2-r^2} )$, where $t$ is the time and
$r=\sqrt{x^2+y^2}$ is the distance of a point from the z-axis, they obtained $O(1,2)$ symmetric solutions using jump boundary conditions applied at the time of collision.  They demonstrated that when the phase
of the Higgs fields is initially different within each
bubble an axial magnetic field forms as the bubbles merge and that
this field has the structure of an expanding ring encircling the
overlap region of the colliding bubbles.

In our earlier work in the MSSM~\cite{stevens1}, EOM were obtained by making an 
expansion about the Kibble-Vilenkin point.  First, $\rho(x)$ was expressed as
\beq
\label{adef}
\rho(x)=\rho_0+a\delta \rho (x)~,
\eeq
with $\rho_0$ the magnitude of the mean scalar field at the center of a single bubble and $a\delta \rho$ fluctuations of the magnitude in the
scalar field once the bubbles merged. Making an
expansion in $a$ as in the Abelian Higgs model we obtained linearized equations within the
bubble overlap region. Collisions in which the Higgs field is relatively unperturbed 
from its mean value when the bubbles merge were termed ``gentle."

Then, assuming as in the Abelian Higgs model that the collision begins at time $t=t_0$ (called $t_c$ in Rf.~\cite{stevens1}), when the bubbles first touch at $z=0$, we used jump boundary conditions to determine the
charged $W^{\pm}$ fields in $O(1,2)$ symmetry and the magnetic field.  These boundary conditions recognize that in some collisions the sign of the z-components of the $W^+$ field at leading order in $a$ is opposite in the two colliding bubbles while the z-components for $W^{-}$ has the same sign, and that for others the phases are the same for $W^\pm$ in the two bubbles. For the first case, referred to with a superscript I, the boundary condition was
\begin{eqnarray}
\label{BCI}
w^{zI}(\tau =t_0,z)&=&w~\epsilon (z) \nonumber \\
\frac{\partial }{\partial\tau} w^{zI}(\tau =t_0,z)&=&0 ~,
\end{eqnarray}
where $\epsilon (z)$ is the sign of $z$, and for the second, identified with a superscript II,
\begin{eqnarray}
\label{BCII}
w^{zII}(\tau =t_0,z)&=&w  \nonumber \\
\frac{\partial }{\partial\tau} w^{zII}(\tau =t_0,z) &=&0 ~.
\end{eqnarray}
Comparing to the Abelian Higgs model we found the two magnetic fields to be of similar size.

\subsection{Effects of the medium:  bubble surface motion and conductivity }
\label{medef}

Because effects associated with the surface break $O(1,2)$ symmetry, a theory formulated with this symmetry and the associated $(z,\tau )$ variables such as that of Ref.~\cite{stevens1} is unsuitable for exploring the consequences of surface dynamics. To explore such effects, the theory has to be formulated in $3+1$ dimensions, and the appropriate symmetry group is $O(3)$.

Additional drawbacks of $O(1,2)$ symmetry include the restriction that $v_{wall}=c$ and difficulty including electrical conductivity in Maxwell's equations. Values of $v_{wall}<1$ have been obtained in diverse studies including modeling collisions of bubbles with constituents of the plasma~\cite{dyne} and solving EOM including nonlinear terms based on an MSSM Lagrangian~\cite{hjk1}. 

A value $v_{wall}\ne 1$ and finite conductivity both directly affect the seed field. This has been discussed comprehensively and their effect estimated in Refs.~\cite{kv95,ae98,cst00} within the context of the Abelian Higgs model.  It was found there that finite conductivity would lead to the decay of the currents (and therefore the magnetic field) with a characteristic time $t_d\approx \sigma/m$~\cite{kv95} with $m$ the gauge boson mass. An additional consequence of the large conductivity arises as follows.  Since the magnetic fields propagate with the speed
of light, for slowly expanding
bubbles these fields would very quickly escape from the region of the
bubble collision in the absence of conductivity. However, because of the large conductivity the magnetic fields become
"frozen" or confined to the region of the bubbles, hindering the escape of magnetic flux into the surrounding false vacuum.  Kibble and Vilenkin showed that the loss of flux is
negligible provided that $\sigma R_c v >>1$, where $R_c$ is the
bubble radius at collision time. With values of conductivity that
are believed to characterize the plasma, currents and magnetic fields
persist on time scales that are long compared to those of the
symmetry breaking scale.

\section{Equations of Motion in the MSSM with $O(3)$ Symmetry }
\label{s:gentle}

In this section we develop a general framework in $O(3)$ symmetry extending our earlier work in Refs.~\cite{stevens1,stevens3a} based on 
the MSSM. Our formulation is intended to be capable of following the 
evolution of bubbles with a given wall speed $v_{wall}$ in $3+1$ 
dimensions, starting at time of nucleation, and determining the magnetic field generated in collisions including effects of finite conductivity. 

We begin the development of our theory in Sect.~\ref{ss:expand} by extending the concept of a gentle collision to the case where the surface is explicitly considered.  We then derive, in Sect.~\ref{ss:formu}, EOM by making an expansion of the scalar field for a pair of bubbles about the mean scalar field. The expansion, justified for gentle collisions, leads to linearized EOM, thus 
simplifying the theory. In Sect~\ref{ss:surface} we discuss some of the new issues that are encountered in solving the EOM when the bubbles are initially separated. As
shown in Sect.~\ref{ss:Maxwell}, the same expansion leads to an expression for the {\it em} current in terms of the $W$ and to a corresponding
Maxwell equation.

\subsection{Gentle collisions in electroweak theory }
\label{ss:expand}

When the surface is considered, we generalize Eq.~(\ref{adef}) by writing
\beq
\label{rhobar}
\rho(x)=  \bar \rho(x) +a\delta \rho (x),
\eeq
with $\bar \rho(x)$ a simple function approximating the mean scalar field at any point $x$ in the medium in the collision.  The quantity $a\delta\rho (x)$ is, as above, the change of the magnitude 
in the mean scalar field induced by the collision. 
In this paper we will obtain, as in Ref.~\cite{stevens1}, linear approximations to the exact non-linear EOM by expanding them in terms of the parameter $a$ appearing in Eq.~(\ref{rhobar}). The resulting EOM are similar to those of Ref.~\cite{stevens1}, but because we now have the surface to consider they differ in a number of essential ways and require the development of completely new techniques to solve.  

The justification of the expansion in terms of $a$ in the present case arises as follows.
Clearly, when two colliding
bubbles are completely separated $\bar \rho(x)$ for these bubbles is, to a very good approximation, the sum of the scalar fields of independent bubbles, and there are no significant fluctuations that need to be considered ($a\approx 0$) to the extent that one has confidence in the choice made for the scalar field in an individual bubble. Additionally, for two completely interpenetrating bubbles $\bar \rho(x)$ within the central region is approximately the same as the scalar field at the center of a single one of the colliding bubbles, and the justification of the expansion is the same there as it was in Ref.~\cite{stevens1}. 

The new issue is to justify the expansion in the peripheral region when two bubbles first begin to merge. The critical point to recognize is that in this region the size of $a$ and hence the accuracy of the expansion will depend on how well the mean field $\bar \rho(x)$ that appears in the EOM approximates the exact scalar field for the colliding bubbles. We come back to this issue below.

\subsection{EOM in electroweak theory for gentle collisions }
\label{ss:formu}

The fact that $\psi$ and $W^d$ (for $d=(1,2)$) enter quadratically in
the $\rho$-equation (Eq.~(8) of Ref.~\cite{stevens1}) places two important constraints on
these quantities:  (1) $\psi$ and $W^d$ must have an expansion in odd
powers of $a^{1/2}$, if we
require the square of these quantities be analytic in $a$; and, (2)
expanding this equation to leading order in $a^{1/2}$,
we find that the terms $\psi^{(0)}$, $w^{(0)1}$, and $w^{(0)2}$ must
vanish.  This is most easily seen in the Euclidean
metric, from the fact that the square of each enters with the same
sign.  However, the same must be true in the Minkowski
metric as well by analytic continuation.  In view of these
considerations, $\psi_\nu$ and
$W^d_\nu$ for $d=(1,2)$ have the following expansion
\begin{eqnarray}
\label{expand1}
\psi_\nu(x)=a^{1/2}\psi_\nu^{(1)}+a^{3/2}\psi^{(3)}_\nu +~...
\end{eqnarray}
\begin{eqnarray}
\label{expand2}
W^d_\nu=a^{1/2}w^{(1)d}_\nu+a^{3/2}w^{(3)d}_\nu +~...~.
\end{eqnarray}
It is natural that an expansion in the same parameter $a^{1/2}$
remains appropriate for $d=3$.  However,
there is no requirement that the leading term vanish, so we take
\begin{eqnarray}
\label{W3}
W^3_\nu =w^{(0)3}_\nu+a^{1/2}w^{(1)3}_\nu+a^{3/2}w^{(3)3}_\nu +~...~.
\end{eqnarray}

With $a\approx 0$ we may take $\bar\rho(x)\approx \rho(x)$, and the $B$-, $\Theta$-, and $W$-equations then give, to first order in $a^{1/2}$,
\begin{eqnarray}
\label{eompsi1}
[\partial^2+\frac{ { \rho}(x)^2}{2}(g^2+g'^2)]\psi^{(1)}_\alpha=0\\
\label{eompsi2}
\partial^\alpha\psi^{(1)}_\alpha=0
\end{eqnarray}
where now
\begin{eqnarray}
\label{psi3}
\psi^{(1)}_\alpha(x)=\partial_\alpha\Theta-\frac{(g^2+g'^2)^{1/2}}{2}Z^{(1)}_\alpha~.
\end{eqnarray}

Equations for $w^{(1)d}_\nu$ may be obtained by expanding the $B$-
and $W$-equations through order $a^{1/2}$.
For $d=$ 1 or 2 (corresponding to $d^{\prime}=$ 2 or 1,
respectively), we obtain
the pair of equations
\begin{eqnarray}
\label{eomu12}
0&=&\partial^2w^{(1)d}_\nu-\partial_\nu\partial\cdot w^{(1)d}+
m(x)^2w^{(1)d}_\nu \nonumber \\
&-&2[\partial^\mu(w^{(0)3}_\nu w^{(1)d^{\prime}}_\mu
-w^{(1)d^{\prime}}_\nu w^{(0)3}_\mu) \nonumber \\
&+&(w^{(1)d^{\prime}}_\mu\partial^\mu w^{(0)3}_\nu
-w^{(0)3}_\mu\partial^\mu w^{(1)d^{\prime}}_\nu) \nonumber \\
& - &(w^{(1)d^{\prime}}_\mu\partial_\nu w^{(0)\mu 3}
-w^{(0)3}_\mu\partial_\nu w^{(1)\mu d^{\prime}})]\nonumber \\
&-&4[(w^{(0)3})^2w^{(1)d}_\nu-w^{(0)3}\cdot w^{(1)d}w^{(0)3}_\nu] ~,
\end{eqnarray}
where $m(x)$ is the mass of the $W$ field,
\beq
\label{wmass}
m(x)^2= \frac{ { \rho}(x)^2g^2}{2}~.
\eeq
The corresponding equation determining $w^{(1)d}_\nu$ for $d=3$ is
\begin{eqnarray}
\label{eomu31}
\partial^2w^{(1)3}_\nu-\partial_\nu\partial\cdot
w^{(1)3}= { \rho}(x)^2g\psi^{(1)}_\nu~,
\label{defu}
\end{eqnarray}
which can be solved once the driving term
$\psi^{(1)}(x)$ has been independently determined from the solution of
Eqs.~(\ref{eompsi1},\ref{eompsi2}).  The considerations for fixing the boundary conditions for $w_\nu^{(1)d}$ and $w_\nu^{(1)3}$ are similar and discussed below. 

This field $w^{(0)3}_\nu$ appearing in Eq.~(\ref{eomu12}) is found to be the
solution of
\begin{eqnarray}
\label{eomu30}
\partial^2w^{(0)3}_\nu-\partial_\nu\partial\cdot w^{(0)3}=0~.
\end{eqnarray}
Because no mass appears in this equation, $W$ occupying this mode propagate at the speed of light and experience no interaction with the scalar field of the bubble to lowest order in $a$, unlike the $W$ described by $w^{(1)d}_\nu$. Because of this, there is no appreciable coupling to $w^{(0)3}_\nu$, and Eq.~(\ref{eomu12}) becomes 
\beq
\label{eomus12}
0=\partial^2w^a_\nu-\partial_\nu\partial\cdot w^a
+m(x)^2 w^a_\nu ~.
\eeq
We see that for sufficiently gentle collisions, {\it all}
relevant equations are linear in $W$.

Simplifying the non-linear $\rho$ equation (Eq.~(8) of Ref.~\cite{stevens1}) using the fact that $w^a_\nu$ and $\psi_\nu$ in leading order go as
$a^{1/2}$, we find that to leading order in $a$
$ \rho(x)$ satisfies the equation
\beq
\label{eomflead2}
0 & = & \partial^2 {\rho} (x)+ { \rho }(x)\frac{\partial V}
{\partial  {\rho}^2}~,
\eeq
and the solution of this equation is clearly identified with $\bar \rho(x)$ appearing in Eq.(\ref{rhobar}).  Methods for solving Eq.~(\ref{eomflead2}) are
discussed in many places, for example Ref.~\cite{kolb}.

In $O(3)$ the complete set of EOM for describing bubble collisions with the surface considered has now been derived and consists of coupled partial differential equations for the relevant fields. Strictly speaking, the set is non-linear because the solution of Eq.~(\ref{eomflead2}) is coupled to the $W$ field through the mass of the $W$, evident in Eq.~(\ref{wmass}). 

However, because Eq.~(\ref{eomflead2}) does not depend on $w^a_\nu$, the coupling of the magnitude of the scalar field ${\rho}(x)$ to the charged $W$ fields $w^{\pm 1}_\nu(x)$ in Eqs.~(\ref{eomus12},\ref{eomflead2}) is particularly simple and allows ${\bar \rho}(x)$ for a system of colliding bubbles to be determined once and for all. Equation~(\ref{eomus12}) is then effectively uncoupled from Eq.~(\ref{eomflead2}) and may be solved directly to obtain $w^a_\nu(x)$ for all $x$, effectively linearizing the EOM and resulting in an enormous simplification.  That the EOM are effectively linear implies that for gentile collisions the coupling of the $W$ to the Higgs dominates the self-coupling of the $W$ fields.  

\subsection{Surface effects and $W^\pm$ fields in bubbles }
\label{ss:surface}

With the solution of Eq.~(\ref{eomflead2})
nearly constant at $\rho(x)=\rho_0$ in the broken phase comprising the interior region of single or overlapping bubbles, our previous 
work~\cite{stevens1} was simplified. However, now that we are considering as well the surface, where $\rho(x)$ begins dropping to
its value in the symmetric phase $\rho(x)=0$ outside we cannot ignore the spatial dependence of the mass as given in Eq.~(\ref{wmass}). The spatial dependence in the surface not only introduces a few technical 
challenges but also, as we will see, new physics with quantitative significance not present in~\cite{stevens1}. 

One of the consequences of the spatial dependence of the $W$ mass can 
be seen by taking the four-divergence of
Eq.~(\ref{eomus12}).  By so doing, we obtain the {\it auxiliary condition}
\beq
\label{aux1}
\chi^a(x)=0~,
\eeq
where
\beq
\label{chidef}
\chi^a(x)\equiv \frac{\partial^\mu( m^2 w_\mu^a)}{m^2}= \partial\cdot
w^a+\frac{w^a\cdot \partial m^2}{m^2}~.
\eeq
Equations~(\ref{aux1},\ref{chidef}) require
\beq
\label{aux2}
\partial\cdot w^a= - \frac{w^a\cdot \partial m^2}{m^2}~.
\eeq
Thus, in contrast to the calculation in Ref.~\cite{stevens1}, it is
no longer true that $\partial\cdot w^a=0$, and as a consequence we
find that the
equations of motion for the $W$ fields more complicated.

The physics becomes clearer by using the relationship in 
Eq.~(\ref{aux2}) to rewrite the EOM in Eq.~({\ref{eomus12}) as
\beq
\label{eomus12a}
0=\partial^2w^a_\nu+ \partial_\nu \frac{w^a\cdot \partial m^2}{m^2}
+m^2w^a_\nu~.
\eeq
The solution to this set of equations is equivalent to the set in 
Eq.~(\ref{eomus12}) provided the
auxiliary condition Eq.~(\ref{aux1}) is
maintained for all $(t,\vec x)$. 

The transformed EOM Eq.~(\ref{eomus12a}) reveal that the spatial dependence of the $W^\pm$ mass provides a perhaps unexpected sensitivity to the bubble surface. The sensitivity occurs through the term $w^a\cdot \partial m^2/m^2$, which becomes in fact divergent in the limit of an infinitely sharp bubble surface.  At this point one cannot rule out significant modifications to results obtained in $O(1,2)$, where the surface is ignored.

To see how the
auxiliary condition Eq.~(\ref{aux1}) may be
maintained for all $(t,\vec x)$, note that Eq.~(\ref{eomus12a}) 
requires $\chi^a(x)$ to satisfy the
Klein-Gordon equation
\beq
\label{kge}
\partial^2 \chi^a(x) + m(x)^2\chi^a(x)=0.
\eeq
By choosing the initial
configuration of $w^a(x)$, at time $t=t_0$, to satisfy
\beq
\label{bc1}
\chi^a(t_0,\vec x)=0
\eeq
and
\beq
\label{bc2}
\frac{\partial \chi^a(t_0,\vec x)}{\partial t}=0~,
\eeq
we assure that $\chi^a(x)=0$ for all future times since Eqs.~({\ref{bc1},\ref{bc2}) are boundary 
conditions for the  
trivial solution $\chi^a(t,\vec x)=0$ of Eq.~(\ref{kge}). Thus,
Eqs.~(\ref{bc1},\ref{bc2}) provide constraints on the initial conditions for the 
$W^{\pm}$ fields.  

To establish the initial conditions requires the choice of a time $t_0$ at which the initial values of the $W$ fields in the bubble are specified. In Ref.~\cite{stevens1} the counterpart of $t_0$ was the point of first contact of the bubbles.  In the current approach $t_0$ may be in fact much earlier, in particular it could be as early as the time of nucleation $t_n$.  
The choice of initial conditions is further discussed in the context of a our model in Sect.~\ref{ss:initcond} below.

These observations make it natural to distinguish two categories of initial conditions when the surface effects are considered.  The first, which we will refer to as boundary conditions, consists of the initial $W$ fields that may be chosen freely.  The second consists of the set determined by Eq.~(\ref{bc1},\ref{bc2}), which we will refer to as the constrained initial conditions.

The definition of $\chi^a$ given in Eq.~(\ref{chidef})
requires $m(x)^2>0$ everywhere as it would be in the mean-field approximation adopted in Sect.~\ref{ss:medium}.

\subsection{Maxwell's equations}
\label{ss:Maxwell}
We may find the Maxwell equation for the {\it em} field
$A^{{\it em}}_\nu(x)$ by taking the linear combination of the $W^{(3)}$
and $B$ indicated in Eq~(\ref{AZ}).  An expression for the corresponding {\it em} current $j_\nu(x)$ consisting of terms quadratic and
cubic in the three fields $W^i(x)$ immediately follows~\cite{stevens1}. 

The result for $j_\nu(x)$ may also be simplified by expanding the
$A^{{\it em}}$ and $W$ fields in powers of $a^{1/2}$.  Letting $a^{(n)}_\nu
(x)$ refer to the terms in the expansion of $A^{{\it em}}_\nu(x)$, we find
that to leading order $A^{{\it em}}_\nu(x) = a^{(2)}_\nu(x)$ and
satisfies the following Maxwell equation,
\begin{eqnarray}
\label{maxwell}
\partial^2a^{(2)}_\nu &-&\partial_\nu\partial\cdot a^{(2)} \nonumber \\
&=& G\epsilon^{ab3} (w_\nu^{(1)b}\partial\cdot
w^{(1)a} \nonumber \\
&-&w^{(1)a}_\mu\partial_\nu w^{(1)\mu b}+2w^{(1)a} \cdot \partial
w^{(1)b}_\nu ) \nonumber \\ &\equiv & 4\pi j_\nu^{\it em} (x)~,
\end{eqnarray}
where we have introduced the coupling parameter $G$ defined in Eq.~(\ref{capG}).  From this we learn that the first non-vanishing contribution to the
{\it em} current is of order $a^{3/2}$ and that it depends on the
components $w^{(1)i}_\nu$ of the {\it charged} $W$ fields ($i=$ 1 and
2), calculated at order $a^{1/2}$.  Expressing the current in terms
of $G$, we find
\begin{eqnarray}
\label{current}
4\pi j_\nu^{\it em} (x)&=&
G\epsilon^{ab3}(w^{(1)b}_\nu \partial\cdot
w^{(1)a} \nonumber \\
&-&w^{(1)a}_\mu\partial_\nu w^{(1)\mu b}+2w^{(1)a}\cdot \partial w^{(1)b}_\nu)~.
\end{eqnarray}
It is easy to prove that this current is conserved,
\beq
\label{consj}
    \partial\cdot j^{\it em}(x)=0
\eeq
using the fact that at the classical level $[w^a_\mu (x)$,$w^b_\nu (x')]=0$ and the fact that the $W$ fields appearing in
Eq.~(\ref{current}) satisfy the EOM, Eq.~(\ref{eomus12}).

So far the bubble has been considered to consist purely of the scalar Higgs field and the associated cloud of charged $W$ gauge bosons coupled to it.
Obtaining the contribution of the charged gauge $W^{\pm}$ fields, Eq.~(\ref{current}), to the {\it em} current for collisions of such bubbles is one of the important results of this derivation. 

However, fermions also contribute to the {\it em} current and have a significant impact on magnetic seed field production.  One contribution was discussed recently in Ref.~\cite{hjk1} and estimated there for the nucleation phase of the collision. Another, discussed in Sect.~\ref{medef}, occurs through the conductivity of the medium $\sigma$. This is one of the most important and best-known contributions and may be taken into account through its associated current $\vec j_c(x)$, 
\beq
\label{concur}
\vec j_c(x)= \sigma \vec E(x)~,
\eeq
where the usual assumption that $\vec j_c(x)$ is proportional to the electric field $\vec E$ has been made.
Detailed calculations of $\sigma$ in the early universe are available~\cite{baym}.

To find Maxwell's equation for the magnetic field $\vec B$, 
\begin{eqnarray}
\label{Bfieldz}
\vec B&=& \vec \nabla\times \vec A^{em}~,
\end{eqnarray}
we multiply Eq.~(\ref{maxwell}) by $\epsilon_{ijk} \partial_j$, obtaining
\beq
\label{max1}
\epsilon_{ijk} \partial_j \partial^2 A_k^{em}&-&\epsilon_{ijk} \partial_j \partial_k \partial\cdot A^{em} \nonumber \\
&=&\epsilon_{ijk} \partial_j j_k^{em} ~.
\eeq
Expresing Eq.~(\ref{Bfieldz}) in components, 
\begin{eqnarray}
\label{Bfield1}
B_i&=& \epsilon_{ijk} \partial_j  A^{em}_k,
\end{eqnarray}
we immediately find the desired result,
\beq
\label{bj}
\partial^2 
\vec B  &=& {\vec \partial }\times {\vec j}^{em}~.
\eeq

\section{Modeling Bubble Collisions in  $O(3)$ Symmetry}
\label{s:solution}

We will assess the importance of surface effects by making numerical simulations that can be meaningfully compared to earlier work in Refs.~\cite{stevens1,stevens3a}. The common dynamical framework is summarized in Sect.~\ref{ss:ourmodel}, with common geometrical aspects specified in Sect.~\ref{ss:colkin}.  The representation of the mean scalar field for two bubbles $\bar \rho(x)=\rho_2(x)$ in this geometry, along with the arguments for its choice, is given in Sect.~\ref{ss:param}. The presence of other bubbles are taken into account in a mean-field approximation in \ref{ss:medium}.  Initial conditions on the $W$ fields, including boundary conditions and the constraints imposed by surface geometry, are discussed in Sect.~\ref{ss:dyniss}. We specialize the theory to cylindrical coordinates in Sect.~\ref{ss:cyncor}. The model is applicable to both weak and strong first-order phase transitions and incorporates some important medium effects absent from our former work.

\subsection{The dynamical framework}
\label{ss:ourmodel}

The familiar conceptual features of our MSSM theory have been presented in Sect.~\ref{ss:firstew}. When the Higgs couples to the other fields as it does in the MSSM through the Lagrangian of Eq.~(\ref{L}), strong and highly
non-perturbative couplings arise forming a tightly coupled many-body system as the
phase transition develops. Specific consequences of this were suggested and embodied in 
Ref.~\cite{stevens1}, and these apply as well to the present work. 

One of the identified consequences of this coupling  
is that as the bubble growth occurs, 
the $W^\pm$ (and other constituents that we are ignoring here) that enter the bubble from the plasma 
gain their mass at the expense of thermal 
energy, a cooling process that continues as volume available to the 
$W$ increases with the bubble expansion. 
Another is that the coupled
fields tend to follow the evolution of the Higgs field {\it coherently}. As they lose thermal energy, the $W$ passing into the bubble enter a 
single mode, a solution to the EOM discussed in the previous section. 

This mode plays a special role, and it is
quite different from the more familiar incoherent thermal modes outside the bubbles. One may think of it as
a coherent state, much like a state of electrons in a superconductor
(except that the $W$ are bosons). The mode of course evolves in time 
according to the EOM, and as the bubble expands and displaces more of 
the volume of the plasma the occupation
of this mode also grows. 
The dynamics driving it is clearly a
non-equilibrium component of the phase transition, and it is a basic assumption of our EOM approach that these coherent fields give rise to the seed fields as bubbles merge before thermal equilibrium is re-established.

There are of course many such field configurations that satisfy the 
EOM, and the one that is realized in a given bubble in the phase 
transition depends on the overall history of the process just 
discussed.  
To calculate the net magnetic field produced in bubble collisions properly, one 
would have to evaluate the field corresponding to each possible initial configuration and average over 
the ensemble of configurations.  

The net effect of this averaging procedure was examined in Ref.~\cite{stevens1}  and found to be  factor less than, but the order of, unity. 
Thus, to get a fair estimate of the net magnetic field it is not necessary to explore the full range of 
possible initial conditions; rather, it is sufficient to examine one initial condition characteristic of the entire ensemble of possibilities.

\subsection{Bubble collision geometry}
\label{ss:colkin}

For the application of our theory to the collision of two bubbles, we will be assuming, as in Ref.~\cite{stevens3a}, 
that the bubbles are nucleated simultaneously at time $t=t_n$ at points $z=\pm z_0$ located symmetrically about the origin on the z-axis with nucleation radii $r_{ns}$. Additionally, here we assume that the bubbles are well separated and non-overlapping before they collide. 

The scalar field of a single bubble may be represented in general as
\beq
\label{rhoform}
\rho(t,\vec x)=\rho_cf^s(t,\vec x)~,
\eeq
where $f^s(t,\vec x)$ is the shape of the field and $\rho_c$ is its magnitude at the center. 
The $f^s(t,\vec x)$ that we will use in this paper, essentially equivalent to that in Ref.~\cite{stevens3a}, is defined in Appendix~\ref{a:scalarpar} in terms the bubble nucleation time $t_n$, nucleation radius $r_{ns}$, surface speed $v_{wall}$, and surface diffuseness $a_s$.  The surface diffuseness is approximately half the distance over which the scalar field falls from its 10\% to 90\% at $t_n$ for large bubbles, {\it i.e.} bubbles with $r_{ns}\gtrsim 2a_s$. 

The bubble nucleated at $z=z_0>0$ is referred to as the right-hand (R) bubble and the one nucleated at $z=-z_0<0$ as the left-hand (L) bubble. We assume axial symmetry so that the relevant spatial coordinates $(r,z)$ for a point are its axial coordinate $r=\sqrt{x^2+y^2}$, its distance from the $z$-axis, and its longitudinal coordinate $z$, its distance from $y-z$ plane that passes through the origin at $z=0$.

It is helpful to think of the collision in terms of the evolution of spatial surfaces that separate regions of the collision occupying true vacuum from those of false vacuum. The connectedness of the surfaces change as the collision process proceeds.
 
Initially, for times $t_n<t<t_c$, where $t_c$ is the collision time or time of merging of the bubbles, the boundary consists of two disconnected surfaces, one for the left-hand bubble $i=L$ and the other for the right-hand bubble $i=R$. For times  $t\geq t_c$ the bubbles coalesce to form a region $i=c$ with a single boundary surface, $S_c$. 

The radius of the bubble surface $R_{1/2}(t)$ is defined as the distance from the center of the bubble (its nucleation point) to the point at which the scalar field has fallen to half its central value. It of course depends on our choice of scalar field, whose details are given in Appendix~\ref{a:scalarpar}. The collision time may then be taken to be the time at which the radius of either bubble becomes equal to the distance of its nucleation point from the origin of the coordinate system, {\it i.e.} when $R_{1/2}(t_c)=z_0$. The solution of this equation for our scalar field is given in Eq.~(\ref{tmfirst}). 

The spatial points forming the boundary surfaces are then determined by the equation  
\beq
\label{boundsurf}
R(r,z)=R_{1/2}(t)~,
\eeq
where
\beq
\label{ridef}
R(r,z)&\equiv& R_R(r,z)\theta(z)+
R_L(r,z)\theta(-z)~, 
\eeq
with
\beq 
\label{RRdef}
R_R(r,z)= \sqrt{r^2+(z-z_0)^2} 
\eeq
being the distance from the center of the right-hand bubble at $z_0$ to the point $(r,z)$ on the surface, and 
\beq
\label{RLdef}
R_L(r,z)= \sqrt{r^2+(z+z_0)^2}
\eeq
being the distance to the same point from the center of the left-hand bubble at $-z_0$. 

After the bubbles merge, the right-hand and left-hand surfaces intersect on the $x-y$ plane in a circle of radius $b(t)=\sqrt{R_{1/2}(t)^2-z_0^2}$ that expands at a rate 
\beq
\frac{db(t)}{dt}= \frac{R_{1/2}(t)}{b(t)}\frac{dR_{1/2}(t)}{dt}~. 
\eeq
Our model of scalar field in the collision of two bubbles, including the region of coalescence is given in Sect.~\ref{ss:param}.

Since the bubble collision geometry is axially symmetric with the relevant spatial coordinates being $(r,z)$, where $z$ is the distance of a point from the $y-z$ plane on the z-axis, and $r=\sqrt{x^2+y^2}$ being its distance from the $z$-axis, cylindrical coordinates is the natural coordinate system for expressing results. In this coordinate system, the $W$ fields may be taken to have the following form,
\beq
\label{wdef}
w^{ a}_\nu (t,\vec r ,z)&=&w^a_0(x),~~\nu=0 \nonumber \\
w^{ a}_\nu (t,\vec r ,z)&=&x_\nu w^a(x),~~\nu=(1,2) \nonumber\\
w^{ a}_\nu (t,\vec r ,z)&=&w^a_z(x),~~\nu=3 ~,
\eeq
with $w^a_z=-w^{z a}$ and $w^a_0=w^{0 a}$. Correspondingly, we write the {\it em} current in cylindrical coordinates as  
\beq
\label{formcurrent}
j_\nu(t,\vec r,z)=(j_0(t,\vec r,z),\vec r j(t,\vec r,z),j_z(t,\vec r,z))~.
\eeq

\subsection{Mean scalar field in two colliding bubbles}
\label{ss:param}

An expression for the mean scalar field $\bar \rho(x)=\rho_2(x)$ for a pair of colliding bubbles may be expressed in terms of the non-overlapping portion of the scalar field, $\Delta\rho(x)$, 
\beq
\label{NO}
\Delta\rho(x)&=& \rho_L(x)(1-\rho_R(x)/\rho_c) \nonumber \\
&&+ \rho_R(x)(1-\rho_L(x)/\rho_c)~,
\eeq
where the scalar fields in the left-hand and right-hand bubbles, while still separated, are taken as
\beq
\label{rhoLRdef}
\rho_{ L}(t,r,z)&=& \rho_c f^s(t,r,z+z_0) \nonumber \\
\rho_{ R}(t,r,z) &=& \rho_cf^s(t,r,z-z_0)
\eeq
with $f^s(t,r,z)$ and $\rho_c$ given in Eq.~(\ref{rhoform}). 

Now, if the scalar 
fields were to simply add
in the overlap region (which we know is not the case~\cite{stevens1}) 
the scalar field would be
\beq
\label{ansatz0}
\rho_L(x)+\rho_R(x)~,
\eeq
which is as large as twice the size of that when they are well separated. What we have learned from the calculation in Ref~\cite{stevens1} is
that the scalar field, when the bubbles have completely merged, is actually half this value, so if we define
\beq
\label{ansatz1}
 \rho_2(x)= \Delta\rho(x)+\frac{1}{2}(\rho_L(x)+\rho_R(x)-\Delta\rho(x))~,
\eeq
then we obtain our {\it ansatz}, which is in agreement with the results
of Ref.~\cite{stevens3a}.

Because there is not much {\it a priori} guidance on how to define $\bar \rho(x)$ in the region where the bubbles begin to overlap, the accuracy of the expansion in terms of $a$ appearing in Eq.~(\ref{rhobar}) is accordingly difficult to establish for any given $\bar \rho(x)$. However, because the peripheral region extends over a small volume relative to that of the two bubbles, it is likely that the corrections for any reasonable choice will be relatively insignificant. 

The evolution of the scalar field in a collision with $\rho_2(x)$ defined in Eq.~(\ref{ansatz1}) is shown in Fig.~\ref{scalar}. This figure shows the scalar field on which the calculations of Sect.~\ref{s:thick} are based. The individual bubbles nucleate at time $t_n=0$ on the $z$-axis centered at $z=\pm z_0$, with $z_0= 35$ and nucleation radius $r_{ns}=20$.  Their scalar fields taken to have wall thickness $a_s=4$ and $v_{wall}=1$. Aside from our choice of surface diffuseness and some compensating adjustment in $r_{ns}$ and $z_0$ (the parameters $z_0$ and $r_{ns}$ are a bit larger than the ones used there so that with the thicker surface the bubbles do not overlap at $t=t_0$) the parameters and collision geometry closely matches that of Ref.~\cite{stevens3a}. Because of the larger surface diffuseness, this corresponds to a somewhat weaker phase transition than that of Ref.~\cite{stevens3a} and thus provides an interesting contrast to the calculation presented there.

The collision time is found from Eq.~(\ref{tmfirst}) to be $t=t_c\approx 11$. The bubble radii at this time are $R_{1/2}(t_c)=z_0=35$.  For times $t<t_c$ the scalar field is approximately confined to two isolated regions, the two individual bubbles, and that after this time to just one region, the collision region. 

\begin{figure}
\centerline{\epsfig{file=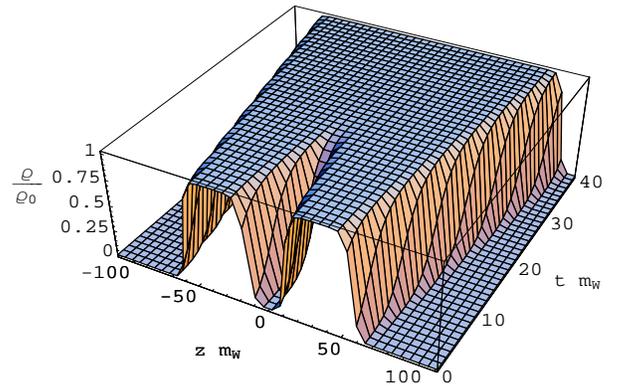,height=6cm,width=8.cm}}
\caption{Evolution of scalar field in the collision of two bubbles.  Nucleation is at $t_n=0$. Field is plotted as a function of $z$ for $r=0$
over the time interval $0\leq t \leq 40$. }
\label{scalar}
\end{figure}

\subsection{Medium containing many bubbles}
\label{ss:medium}

In a medium undergoing a first-order phase transition there are many distinct regions $S_i$ of scalar potential, each of which is either a single bubble or a collection of mutually overlapping bubbles.  The net scalar field $ \rho(x)$ for this system may thus be expressed as a sum over the scalar fields $\rho^{(i)}(x)$ for each of the $N$ regions,
\beq
\label{nono}
 \rho(x)= \sum_i^N \rho^{(i)}(x) ~.
\eeq
For such a system, the mass of a $W^\pm$ boson at any point $x$ continues to be given by Eq.~(\ref{wmass}), where $\rho(x)$ is the net scalar potential at the location of the $W$. 

In this paper we are interested in the evolution of at most two bubbles in this sum.  But in tracking their evolution, we should not ignore the presence of the other bubbles. We will account for them in a mean-field approximation by averaging over their locations and representing their collective effect by an average scalar field $\rho_{av}$, 
\beq
\label{avrho}
\rho_{av}&=& \langle \sum_i^{N^\prime}\rho^{(i)}(x) \rangle
\nonumber \\
&\equiv&\frac{1}{V} \int d^3x \sum_i^{N^\prime}\rho^{(i)}(t,\vec x) ~,
\eeq
where $V$ is the total volume over which the integral runs and the prime on the sum means we exclude the region(s) of explicit interest in the average. 
The quantity $\rho_{av}$ thus acquires its specific value from the presence of other bubbles that appear as the phase transition develops. 

Since $\rho_{av}$ is essentially the average scalar field arising from the bubbles in the medium at the time of interest, at the onset of the phase transition $\rho_{av}\approx 0$ and at the completion of the phase transition the scalar field becomes uniform with $\rho_{av} \to \rho_0$.  Thus, $\rho_{av}/\rho_0$, with  $0<\rho_{av}/\rho_0<1$, is not only a measure of the extent to which the phase transition has evolved but also tracks the relative density of the bubbles in this evolution.
One might expect that the binary collisions that are of most interest in this paper dominate for $\rho_{av}/\rho_0 \lesssim 0.1 - 0.2$ and that for larger values simultaneous collisions of multiple bubbles begin to contribute significantly.

Accordingly, taking into account the average value of the mean scalar field $\rho_{av}$, and as long as $\rho_{av}/\rho_0\lesssim 0.1 - 0.2$, the variation of the mass of a $W$ in the left-hand and right-hand bubbles when they are separated is given by 
\beq
\label{indivb}
m_L(x)^2& =&\frac{ g^2}{2}(\rho_{av} +{\rho}_L(x))^2 \nonumber \\
m_R(x)^2& =&\frac{ g^2}{2} (\rho_{av} +{\rho}_R(x))^2~,
\eeq
respectively.
In general, for a pair of bubbles in a medium experiencing any degree of overlap, $m(x)$ is given by 
\beq
\label{mav3a}
m(x)^2 &=&\frac{g^2}{2}( \rho_2(x)+\rho_{av})^2 ~,
\eeq
where $\rho_2(x)$ is defined in Eq.~(\ref{ansatz1}).

With $m(0)$ referring to the mass at the center of one of the regions $S_i$ of the collision, clearly
\beq
\label{mav3}
m(0)^2 &\equiv&\frac{g^2}{2}\rho_0^2= m_W^2~.
\eeq
Since it is also true from Eq.~(\ref{mav3a}) that 
\beq
\label{mav3b}
m(0)^2 &=&\frac{g^2}{2}( \rho_2(0)+\rho_{av})^2 ~,
\eeq 
and the scalar potential $\rho_2(0)=\rho_c$ it follows that at the center of a bubble in the medium the scalar field is
\beq
\label{mav4}
\rho_c &=& \rho_0 - \rho_{av}~.
\eeq

The quantity $\rho_{av}$ determines the mass of the $W$ in the bubble surface and thus plays a role in confining the $W$ to the region of the bubble.  
Beyond this, the results of the theory are relatively insensitive to the specific value of $\rho_{av}$.

\subsection{Initial conditions on $W$ fields}
\label{ss:dyniss}

A meaningful comparison of the present theory to those of Refs.~\cite{stevens1,stevens3a} clearly requires that similar initial conditions be applied. In contrast to Ref.~\cite{stevens1}, the initial conditions are imposed here on $W$ fields in bubbles that are initially separated.

Although our intent is to assess the importance of surface dynamics by comparing the results of our present paper to those of Refs.~\cite{stevens1,stevens3a}, we should keep in mind that there is no well-defined limit in which the present theory, which exhibits $O(3)$ symmetry, {\it exactly} coincides with that of Ref.~\cite{stevens1}, which exhibits $O(1,2)$ symmetry.

\subsubsection{Boundary conditions for individual bubbles}
\label{sss:chargen}

As indicated earlier, in the dynamical framework of Refs.~\cite{stevens1,stevens3a} as a bubble expands it displaces $W$ in the thermal plasma which enter the bubble in a single mode. The normalization of this mode in the bubble (that is, in the true vacuum) is determined by requiring that the average number density of $W^\pm$ in the bubble be equal to the number density of the $W^\pm$ quanta in the displaced plasma.
This condition maintains the average density of $W^+$ to be roughly constant as a function of time and equal to the density of the $W^-$ inside the isolated bubble. 
As a consequence
\beq
\label{neut}
|w^{a=1}_\mu(t, \vec x )|=|w^{a=2}_\mu(t, \vec x)|
\eeq
with random relative phases.
The linearity of the EOM in Eq.~(\ref{eomus12}) guarantees Eq.~(\ref{neut}) in isolated bubbles if a some initial time $t_0$ $w^a_\mu$ has the same magnitude and shape for both $a=(1,2)$,
\beq
\label{wbnd}
w^{a=1}_\mu(t_0, \vec x)&=&\alpha w^{a=2}_\mu(t_0, \vec x) \nonumber \\
\frac{\partial }{\partial t}w^{a=1}_\mu(t_0, \vec x)&=& \alpha 
\frac{\partial }{\partial t}w^{a=2}_\mu(t_0, \vec x)~,
\eeq
where $\alpha$ is an arbitrary phase.

Because the electromagnetic current in Eq.~(\ref{current}) is antisymmetric in the labels $a$ and $b$ this current will vanish since the field $w^a(x)$ for $a=1$ has the same dependence on $x$ as that for $a=2$.  Thus, in an isolated bubble the electromagnetic current will vanish.

However when two bubbles, each meeting the condition in Eq.~(\ref{neut}) collide, it is in general not the case that $w^a_\mu(x)$ satisfying the equations of motion will be proportional in the region of overlap, and as a consequence {\it em} currents will form.
This is the reason why magnetic fields are in general produced when bubbles collide.  We model this below by solving the equations for the $W$ fields with different sets of boundary conditions on $w_z^a$ for $a=1$ and $a=2$.

\subsubsection{Boundary conditions for bubble collisions}
\label{sss:BCond}

With Eq.~(\ref{wbnd}) in mind, we turn our attention to finding boundary conditions for collisions so that the $W^\pm$ fields are as similar as possible to those of Refs.~\cite{stevens1,stevens3a}. By requiring that the bubbles be separated at the initial time $t=t_0$~\cite{stevens3a} and that Eq.~(\ref{wbnd}) is satisfied there is no {\em} field produced until the collision occurs, as in Refs.~\cite{stevens1,stevens3a}. 

The analog of the jump boundary conditions in Eqs.~(\ref{BCI},\ref{BCII}) at $t_0$ in $O(3)$ are then
\beq
\label{bcI}
w^{I}_z(t_0,r,z) &=&w_{L}(r,z)-w_{R}(r,z)
\nonumber \\
\frac{\partial w^{I}_z(t_0, \vec x) }{\partial t} &=& 
\frac{\partial w_{L}(r,z) }{\partial t} - \frac{\partial
w_{R}(r,z) }{\partial t} ~,
\eeq
and
\beq
\label{bcII}
w^{II}_z(t_0,r,z)&=&w_{L}(r,z)+w_{R}(r,z) \nonumber \\
\frac{\partial w^{II}_z(t_0,\vec x)  }{\partial t} &=& 
\frac{\partial w_{L}(r,z) }{\partial t} + \frac{\partial
w_{R}(r,z)}{\partial t} ~. 
\eeq
The functions $w_{ L}(\vec x)$ and $w_{ R}(\vec x)$ are clearly the 
profiles of the $W$ fields in the left-hand and right-hand bubbles, respectively, at 
time $t=t_0$. To coincide with Eqs.~(\ref{BCI},\ref{BCII}), the sign of $w^{a=1}_z$ has been chosen ${\it opposite}$ to that of $w^{a=2}_z$ in one of the bubbles, while $w^{a=1}_z$ and $w^{a=2}_z$ have been chosen to have the same sign in the other. The reason for the choice of the initial time derivatives is explained below. 

As discussed at the end of Sect.~\ref{sss:chargen}, at $t=t_0$ and the functions $w_{ L}(\vec x)$ and $w_{ R}(\vec x)$ have the same magnitudes and shapes. Thus we write 
\beq
\label{wLRdef}
w_{ L}(\vec x)&=& n_Wf^{w}(t_0,r,z+z_0) \nonumber \\
w_{ R}(\vec x) &=& n_Wf^{w}(t_0,r,z-z_0)~,
\eeq
where $f^w(t_0,r,z\pm z_0)$ chosen similar to $f^s(t_0,r,z\pm z_0)$ of the scalar field so that the $W$ fields fill the bubbles uniformly~\cite{stevens1}. The normalization constant $n_W$ is determined as in Refs.~\cite{stevens1,stevens3a} allowing
$W^d_\nu$ in Eq.~(\ref{expand2}) to be identified with $w^{(1)d}_\nu$ rather than with $a^{1/2}w^{(1)d}_\nu$. 
Of course for times $t>t_0$ the distribution of the $W$ in the bubble will differ from that of the scalar field since these fields evolve through different EOM.

Boundary conditions on all the fields $w^a_\nu$ and their time derivatives are also required to solve the EOM. According to the requirements of Sect.~\ref{ss:dyniss}, $w^a$ and $w_0^a$ should be empty at $t=t_0$, so the boundary conditions on these quantities are 
\beq
\label{bcc2}
w^{I}(t_0, \vec x )&=& w^{II}(t_0, \vec x )=0 \nonumber \\
w^{I}_0(t_0, \vec x )&=& w^{II}_0(t_0, \vec x )=0~.
\eeq
Boundary conditions on their time derivatives are determined by the constrained initial conditions, discussed below.

The initial time derivatives in Eqs.~(\ref{bcI},\ref{bcII}) assure that $W$ of the plasma displaced by the expanding bubble end up populating the coherent mode inside the bubble in our $O(3)$ formulation. 
To achieve this, we take
\beq
\label{wzLRderdef}
\frac{\partial w_{L}(x,y,z) }{\partial t} &=&
n_W'\frac{\partial f^w(t_0, x,y,z+z_0) }{\partial t} \nonumber \\
\frac{\partial w_{R}(r,z) }{\partial t} &=&
n_W'\frac{\partial f^w(t_0, x,y,z-z_0) }{\partial t} ~,
\eeq 
where the choice of $n_W'$ is discussed in Appendix~\ref{a:bcderwz}. 
As long as $v_{wall}>v_W$, $n_W'$ is determined by $f^w$ and $f^s$ according to Eq.~(\ref{nwprime})\footnote{ Note that the speed of a $W$ boson in the surface of the bubble is given by $v_W\approx p_W/\sqrt{p_W^2+ {\bar m^2_W }}$ where $p_W\sim \hbar/R_{0s}(t_0)$ is determined by the zero point motion of the $W$ in the bubble and $\bar m_W$ is the $W$ mass in the surface of bubble.  For cases of interest $v_W$ is expected to be less than $v_{wall}$.}. In general, $n_W'\neq n_W$, and the simple and natural choice $f^w=f^s$ at $t=t_0$ gives $n_W=n_W'$. Perhaps unexpectedly, with the motion of the surface taken into account in our $O(3)$ formulation, the analog of the boundary condition on the derivative of the $w_z$ field in Eqs.~(\ref{BCI},\ref{BCII}) of the $O(1,2)$ formulation that maintains the average density of the $W$ is not $n_W'=0$. 
 
With the motion of the surface taken into account in this fashion, the boundary condition on the derivative of the field continues to be the mechanism by which the average density of the $W$ remains constant inside the bubble and expands with it. This happens in the $O(1,2)$ formulation of Ref.~\cite{stevens1} as a consequence of the boundary condition $\partial w_z(\tau_0,z)/\partial \tau=0$.

Choosing the distribution in Eq.~(\ref{wLRdef}) to have the shape of the scalar field $f^s(t,r,z)$ in Fig~\ref{scalar}, the boundary conditions $w_z^{I}(t_0,r,z)$ and $w_z^{II}(t_0,r,z)$ are shown in Figs.~\ref{f:bcIwz} and~\ref{f:bcIIwz} (with $n_W=1$). We will see explicitly how the {\it em} currents and associated magnetic fields begin to form in the collision once the bubbles begin to overlap when we solve the EOM in Sect.~\ref{s:thick}. 

For these calculations, we find it convenient to normalize the $w_z^a$ fields to unity inside the bubbles rather than to $n_W$ and $n_W'$ as in Eqs.~(\ref{wLRdef},\ref{wzLRderdef}). The normalization re-enters the calculation as a factor in the normalization of the {\it em} current as found in Sect.~\ref{ss:cyncor}.

\begin{figure}
\centerline{\epsfig{file=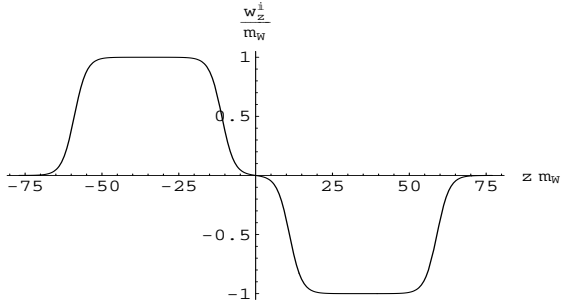,height=4cm,width=8.cm}}
\caption{ The initial condition $w_z^{I}(t_0,r,z)$ for $r=0$. }
\label{f:bcIwz}
\end{figure}

\begin{figure}
\centerline{\epsfig{file=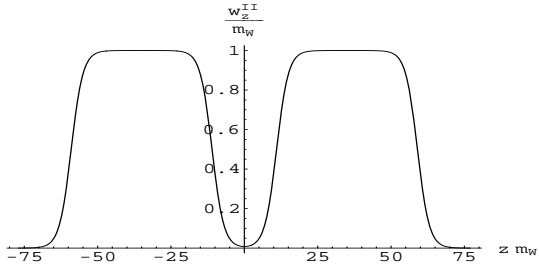,height=3.5cm,width=8.cm}}
\caption{ The initial condition $w_z^{II}(t_0,r,z)$ for $r=0$. }
\label{f:bcIIwz}
\end{figure}

\subsubsection{Auxiliary condition and constrained initial conditions}
\label{ss:initcond}

We have used the freedom in choosing the boundary conditions to match our calculation to that in Refs.~\cite{stevens1,stevens3a}. However, as discussed earlier, the appearance of a spatial-dependent mass $m(x)$ means that not all of the initial conditions are independent and must satisfy the constraints in Eqs.~(\ref{bc1},\ref{bc2}). Although this is a complication that was not a significant issue in Ref.~\cite{stevens1}, in $O(3)$ symmetry they are easily maintained in cylindrical geometry, which is used for obtaining numerical results in subsequent sections.  We will use these conditions to fix the time derivatives of $w^a$ and $w_0^a$ at $t=t_0$ using results found in Appendix~\ref{a:initcondA}.

\subsection{Cylindrical coordinates}
\label{ss:cyncor}

In view of the axial symmetry of the collision as specified in Sect.~\ref{ss:colkin} with cylindrical coordinates the natural coordinate system for expressing results, the EOM in Eq.~({\ref{eomus12a}) with $w^a_\nu(x)$ given in Eq.~(\ref{wdef}) and {\it em} currents in Eq.~(\ref{current}) with $j_\nu(x)$ given in Eq.~(\ref{formcurrent}) are expressed in cylindrical coordinates in Ref~\cite{stevens3a}.  For convenience, they are reproduced here,
\beq
\label{eqw1a}
0&=&(\frac{\partial^2}{\partial t^2} -\frac{\partial^2}{\partial
r^2}-\frac{1}{r}\frac{\partial}{\partial r}
-\frac{\partial^2}{\partial z^2} +m^2) w^{a}_0 +2\nonumber \\
    &\times&\frac{\partial}{\partial t} ( w_0^a \frac{\partial \ln m}{\partial t}
+rw^a \frac{\partial \ln m}{\partial r} 
- w^a_z\frac{\partial \ln m}{\partial z} )~,
\eeq
\beq
\label{eqw2a}
0&=&(\frac{\partial^2}{\partial t^2} -\frac{\partial^2}{\partial
r^2}-\frac{3}{r}\frac{\partial}{\partial r}
-\frac{\partial^2}{\partial z^2} +m^2) w^{a} -\frac{2}{r} \nonumber \\
   &\times& \frac{\partial}{\partial r}  (w_0^a \frac{\partial \ln m}{\partial t}
+rw^a \frac{\partial \ln m}{\partial r}
-w^a_z\frac{\partial \ln m}{\partial z})~,
\eeq
\beq
\label{eqw3a}
0&=&(\frac{\partial^2}{\partial t^2} -\frac{\partial^2}{\partial
r^2}-\frac{1}{r}\frac{\partial}{\partial r}
-\frac{\partial^2}{\partial z^2} +m^2) w^{a}_z +2\nonumber \\
    &\times& \frac{\partial}{\partial z} ( w_0^a \frac{\partial \ln m}{\partial t}
+rw^a \frac{\partial \ln m}{\partial r} 
-w^a_z\frac{\partial \ln m}{\partial z}) ~.
\eeq
These EOM are equivalent to Eqs.~(\ref{eomus12}) when constraints imposed by the
auxiliary condition of Eq.~(\ref{aux1}) are satisfied; see Appendix~\ref{a:initcondA} for the details.  
When these constraints are
satisfied, Eqs.~(\ref{eqw1a},\ref{eqw2a},\ref{eqw3a}) determine the
$W$ fields, $w^a_\nu$.  

Equation~(\ref{formcurrent}) expressed in terms of these $w^a_\nu$  provide the {\it em} currents given in Ref.~\cite{stevens3a}, 
\begin{eqnarray}
\label{j0final2}
4\pi j_0 (x)&=&
G\epsilon^{ab3}( w^{b}_0 \partial\cdot
w^{a} 
+ w^{a}_0 \frac{\partial w^{b}_0 }{\partial t} \nonumber \\
&+&r^2 w^{a}\frac{\partial w^{b}  } {\partial t}  
- w^{a}_z\frac{\partial w^{z b}  }{\partial t}\nonumber \\
&+& 2rw^{a} \frac{\partial w^{b}_0 }{\partial r} 
-2w^{a}_z \frac{\partial w^{b}_0  }{\partial z} )~,
\end{eqnarray}
\beq
\label{jfinal2}
4\pi j(x)&=&
G\epsilon^{ab3}(w^{b}  \partial\cdot w^{a} 
+ rw^{a} \frac{\partial w^{b} }{\partial r} \nonumber \\
&+& \frac{1}{r}  w^{a}_0\frac{\partial w^{ b}_0  }{\partial r} +\frac{1}{r} w^{a}_z\frac{\partial w^{z b}  }{\partial r} \nonumber \\
&+&  2w^{a}_0 \frac{\partial w^{b} }{\partial t} 
- 2w^{a}_z \frac{\partial w^{b} }{\partial z})~,
\end{eqnarray}
\begin{eqnarray}
\label{jzfinal2}
4\pi j_z (x) 
&=&G\epsilon^{ab3}(w^{b}_z \partial\cdot
w^{a} - w^{a}_z \frac{\partial w^{b}_z }{\partial z} \nonumber \\
&-& w^{a}_0\frac{\partial w^{ b}_0  }{\partial z}+r^2 w^{a}\frac{\partial w^{b}  } {\partial z}  \nonumber \\
&+&2  w^{a}_0 \frac{\partial w^{b}_z }{\partial t} + 2rw^{a} \frac{\partial w^{b}_z }{\partial r})  ~.
\end{eqnarray}
With these expressions, current conservation
\beq
\label{currentcons}
\partial_\nu j^\nu(t,\vec r,z)&=& \frac{\partial j_0(t,\vec
r,z)}{\partial t} + \frac{1}{r}\frac{\partial r^2 j(t,\vec
r,z)}{\partial r} \nonumber \\
&-& \frac{\partial j_z(t,\vec r,z)}{\partial z} =0~,
\eeq
can be shown to be satisfied as expected, and Maxwell's equation for $\vec B$, Eq.~(\ref{bj}), gives the desired seed fields. From these results we can infer the importance of surface dynamics for generating magnetic fields in bubble collisions.

Adopting the convention that $w^a_z$ with BCI and BCII are normalized to unity (in units of $m_W$) in the bubble at $t=t_0$, the overall normalization of the {\it em} current
is the product of the factor $G$
and the square of $n_W$ introduced and discussed in Sect.~\ref{sss:BCond}.
Thus
\beq
\label{currentnorm}
G\rightarrow n_W^2 G~, 
\eeq
where $G$ is calculated using the values of $g$ and $g'$ quoted below Eq.~(\ref{AZ}) and the
value of $n_W$ is specified in Eq.~(72) of the Appendix of Ref.~\cite{stevens1}, fixing the average number density of $W$ in the bubble equal to the number density of those $W$ quanta in the thermal plasma that can
make a transition into the bubble without violating energy
conservation.  In this way, we find
that Eq.(\ref{currentnorm}) becomes
\begin{eqnarray}
\label{currentnorm1}
n_W^2 G &\approx& -38.5+1.36\hat T_c~ {\rm GeV} \nonumber \\
&=&2.32 m_W~,
\end{eqnarray}
taking the transition temperature (in GeV) to be $\hat T_c= 166$ from Ref.~\cite{eikr}.

With the {\it em} current having the form given in Eq.~(\ref{formcurrent}), one can show that the solution of Eq.~(\ref{bj}) takes the form
\begin{eqnarray}
\label{Bfield}
\vec B&=& \frac{(-y,x,0)}{r} B^\phi ~,
\end{eqnarray}
where $(-y,x,0)/r$ is a unit vector in the azimuthal direction.  Thus, the
magnetic field generated in the bubble collision lies in the azimuthal plane, with $x$ and $y$ components only, and encircles the z-axis, just as in the Abelian Higgs
model and our earlier work~\cite{stevens1}. 

Using Eqs.~(\ref{bj},\ref{Bfield}) and including the conductivity current from Eq.~(\ref{concur}), we immediately obtain $B^\phi$ in cylindrical coordinates as the solution of
\beq
\label{maxazb1}
&&(\frac{\partial^2}{\partial t^2} -r\frac{\partial^2}{\partial
r^2}\frac{1}{r} -3\frac{\partial}{\partial r} \frac{1}{r}
-\frac{\partial^2}{\partial z^2}+\sigma \frac{\partial}{\partial t} ) B^\phi \nonumber \\
&=&4 \pi j^\phi~,
\eeq
where
\beq
\label{jcdef}
4\pi j^\phi=
4\pi\frac{\partial j_z (x)}{\partial r} + 4\pi r \frac{ \partial j
(x)}{\partial z} ~.
\eeq
and we have used the fact that 
\beq
\vec \partial \times \vec E= -\frac{\partial \vec B}{\partial t}~.
\eeq

In obtaining the above results, the following identities are useful,
\beq
\label{cyn1}
\partial\cdot w^a= \frac{\partial w^{a}_0 }{\partial t}
+2w^a+r\frac{\partial w^a }{\partial r} - \frac{\partial w^a_z
}{\partial z}~,
\eeq
\beq
\label{cyn3}
w^a\cdot\partial &=& w^{a}_0 \frac{\partial }{\partial t}
+rw^a\frac{\partial }{\partial r} - w^a_z \frac{\partial }{\partial z} ~,
\eeq
and
\beq
\label{cyn2}
\frac{w^a\cdot \partial m^2}{m^2}
&=& \frac{w_0^a}{m^2} \frac{\partial m^2}{\partial t}
+\frac{rw^a}{m^2} \frac{\partial m^2}{\partial r}
-\frac{w^a_z}{m^2}\frac{\partial m^2}{\partial z} \nonumber \\
&=& 2( w_0^a \frac{\partial \ln m}{\partial t}
+rw^a \frac{\partial \ln m}{\partial r} \nonumber \\
&-&w^a_z\frac{\partial \ln m}{\partial z} ) ~.
\eeq

\section{Numerical Results, $a_s=4$}
\label{s:thick}

Because the magnetic field is critical for determining whether present day galactic fields could have been seeded during the primordial EWPT, it is the most important prediction of our theory and the focus of our numerical study. In the process of determining it we will explore the 
role played by the bubble surface in the production of these fields, as this has not been examined in previous studies.  
To achieve this understanding we will examine the $W$ fields, the source of the ${\it em}$ currents in our MSSM formulation, as well as the currents themselves.

In this section we assume $v_{wall}=1$.  Although our theory as formulated accommodates wall speeds $v_{wall}<1$ and $\sigma \neq 0$, we postpone discussion of these effects to a later section.

Our scalar field in an isolated bubble for $a_s=4$ is the same as the one in our example of Sect.~\ref{ss:param}. With the scalar field of an isolated bubble chosen in this way, the scalar field describing the collision is the same as the one shown in Fig.~\ref{scalar}. The collision assumed to take place in an average scalar background field of $\rho_{av}=.1m_W$ ($\rho_{av}/\rho_0=0.046$), corresponding to a collision early in the evolution of the phase transition.

With this scalar field, at $t=t_0$ the radius of each bubble is $R_{1/2}(t_0)\approx 24$, so the bubble surfaces are separated on the $z$ axis by $2(z_0-R_{1/2}(t_0))\approx 22$.  Although we show results for $t_0<t<t_c+20$, we have solved the EOM over the longer interval $t_0<t<t_c+\delta t_{max}$, where with $\delta t_{max}=29$, at which time the radius of each bubble has increased to $R_{1/2}(40)\approx 64$. The numerical accuracy of the solution deteriorates rapidly for $\delta t >20$.

We show results for both nucleation and collisions.  The nucleation stage corresponds to times $t<t_c=11$, during which time the bubbles may be considered as evolving approximately independent of one another.  Collisions then correspond to the interval $t_c<t<t_c+20$.

\subsection {Nucleation stage of bubble evolution, $a_s=4$}
\label{s:nucleathick}

During the nucleation stage of evolution, prior to the collision, both colliding bubbles have scalar fields with approximately the same shape and wall speed.  In this case, the $W^a$ fields are proportional for $a=I$ and $a=II$, and it is therefore sufficient to examine the fields in just one of the two bubbles. For this reason, for nucleation we drop the superscript distinguishing the two boundary conditions. To simplify the discussion of nucleation, we calculate the $W$ fields in a coordinate system translated so that nucleation originates at the origin, $z_0=0$.

The boundary conditions for our numerical simulations of nucleation are 
\beq
\label{bcIc11n}
w_0(t_0,\vec x)&=&0 \nonumber\\
w(t_0,\vec x)&=&0 \nonumber \\
w_z(t_0, r,z )&=& f^w(t_0,r,z) \nonumber \\
\frac{\partial }{\partial t} w_z(t_0,r,z)&=&\frac{\partial f^w(t_0,r,z) }{\partial t} ~,
\eeq
where $f^w(t_0,r,z)$ is the same quantity appearing in Eq.~(\ref{wLRdef}).  Because these are the bubbles that initiate the collisions presented in Sect.~\ref{s:collthick} below, the parameters of $f^w(t_0,r,z)$ are clearly the same as those for the collision (with $n_W=n_W'$). Using the boundary conditions of Eq.~(\ref{bcIc11n}), Eq.~(\ref{spec0n}) specifies the initial condition for $\partial w(t_0, \vec x)/\partial t $ and Eq.~(\ref{nuc0t})  the initial condition for $\partial w_0(t_0, \vec x)/\partial t$.
The $W$ fields are then calculated solving the EOM given in Sect.~\ref{ss:cyncor} with boundary conditions specified in Eq.~(\ref{bcIc11n}) with the scalar field of the bubble appearing in the example of Sect.~\ref{ss:param}.
The resulting $W$ fields are shown as a function of $(t,z)$ for $r\approx 0$ in Figs.~\ref{f:wzNmotion},\ref{f:wNmotion},\ref{f:w0Nmotion}. 

It is clear from the results that the $W$ fields undergo a time-dependent evolution from nucleation at $t=t_n=0$ to the time of collision at $t=t_c=11$.  The dominant component of the field is of course $w_z(t,r,z)$, since $w_0$ and $w$ vanish at nucleation by choice of the boundary conditions.  These smaller components of the field grow with time in part because their time derivatives, given by the initial conditions, are finite at $t=t_0$. There is no magnetic field associated with a single bubble during this interval for reasons discussed in Sect.~\ref{sss:chargen}.  However, there is a small field generated prior to $t=t_c$ in the collision of two bubbles as their surfaces, which have a finite diffuseness, begin to interpenetrate.
 
\begin{figure}
\centerline{\epsfig{file=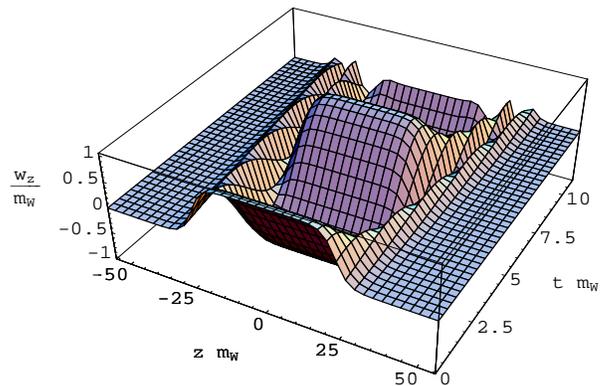,height=6cm,width=8.cm}}
\caption {Evolution of $w_z$ field in a bubble after nucleation. The field is plotted as a function of $(t,z)$ for $r\approx 0$.}
\label{f:wzNmotion}
\end{figure}

\begin{figure}
\centerline{\epsfig{file=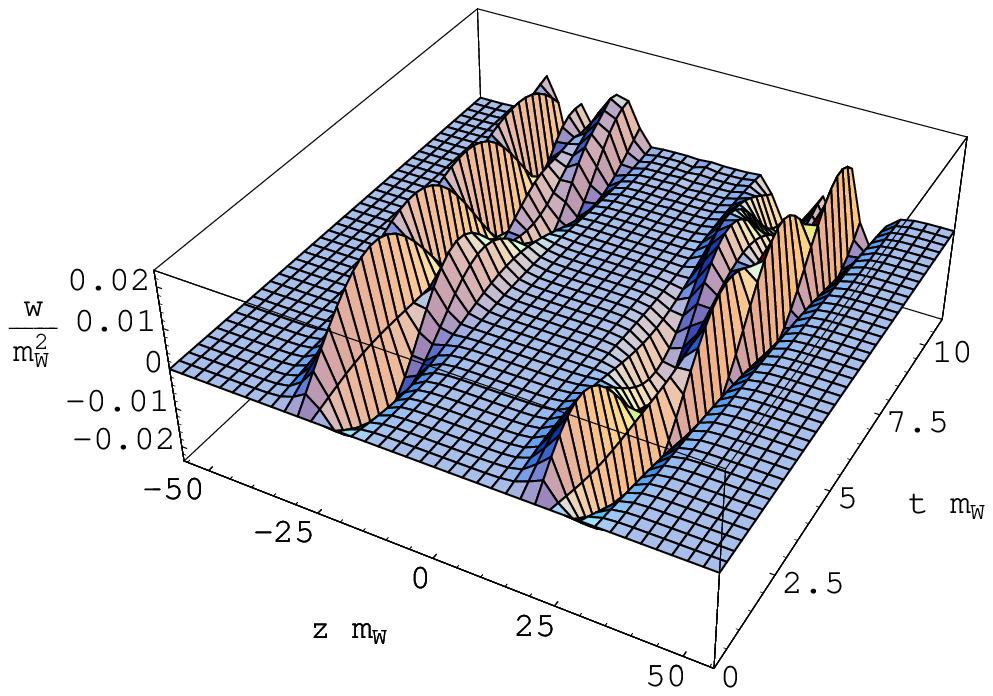,height=6cm,width=8.cm}}
\caption {Evolution of $w$ field in a bubble after nucleation. The field is plotted as a function of $(t,z)$ for $r\approx 0$. }
\label{f:wNmotion}
\end{figure}

\begin{figure}
\centerline{\epsfig{file=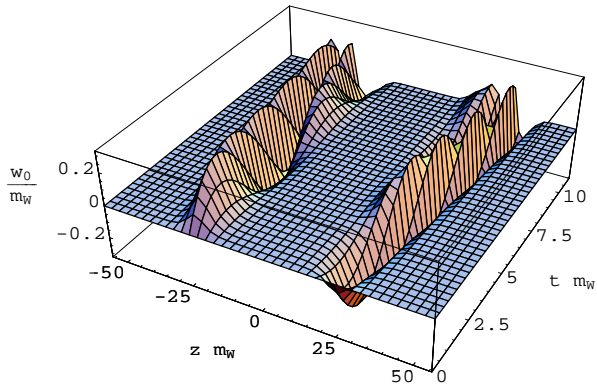,height=6cm,width=8.cm}}
\caption {Evolution of $w_0$ field in an isolated bubble. The field is plotted as a function of $(t,z)$ for $r\approx 0$. }
\label{f:w0Nmotion}
\end{figure}

\subsection {Bubble collisions, $a_s=4$}
\label{s:collthick}

We turn our attention to collisions, the source of magnetic fields. 
Using the boundary conditions given in Eqs.~(\ref{bcI},\ref{bcII}) and shown in Fig.~\ref{f:bcIwz} and Fig.~\ref{f:bcIIwz} and the boundary conditions in Eqs.~(\ref{bcc2}), Eqs.~(\ref{specII},\ref{specI}) give the constrained initial conditions for $\partial w^{II}(t_0, \vec x)/\partial t $ and $\partial w^{I}(t_0, \vec x)/\partial t $.  Likewise, initial conditions for $\partial w_0^{II}(t_0, \vec x)/\partial t$ and $\partial w_0^{I}(t_0, \vec x)/\partial t$ are given in Eqs.~(\ref{bcII0t},\ref{bcI0t}).

With the scalar field describing the collision as shown in Fig.~\ref{scalar}, the profile of the region of bubble overlap in the collision at $t=t_c+20\approx 31$ is shown in Fig.~\ref{f:overlap}. At this time, the collision region extends along the $z$-axis from $-R_{1/2}(31)-z_0=-91 <z<R_{1/2}(31)+z_0=91$.

\begin{figure}
\centerline{\epsfig{file=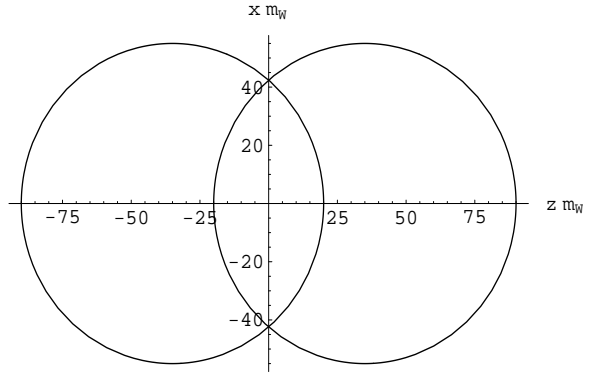,height=5cm,width=7.72cm}}
\caption{Showing the projection of the surface $S_c$ of the collision region onto the $x-z$ plane at time $t_f= t_c+20$ with $v_{wall}=1$. The surfaces $S_c$ is defined by Eq.~(\ref{boundsurf}) with $R_{1/2}(t_f)=55$.}
\label{f:overlap}
\end{figure}
The fields $w_z^a(t,r,z)$, $w_0^a(t,r,z)$, and $w^a(t,r,z)$ for the collision are determined by solving the EOM given in Sect.~\ref{ss:cyncor} on the interval $t_0\leq t\leq t_c+\delta t_{max}$, where $t_0=0$ and $\delta t_{max}=29$. 
The results are illustrated showing $w_z^I$ in Fig.~\ref{f:WzImove} plotted in the $x-z$ plane at $t=t_c+\delta t$ with $\delta t=20$.
The region over which $w_z\neq 0$ delineates the region $i=c$ of bubble coalescence, with the bubble collision region clearly in evidence.  

\begin{figure}
\centerline{\epsfig{file=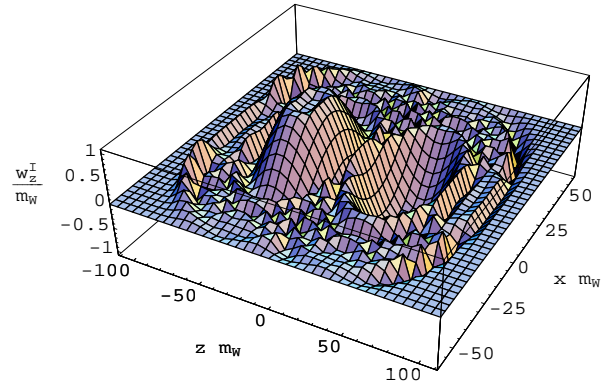,height=6cm,width=8.cm}}
\caption{ $w_z^I(t,\vec x)$ in the $x-z$ plane at $t= t_c+20$ for collisions of bubbles.}
\label{f:WzImove}
\end{figure}

The {\it em} current $j_z(t,r,z)$, $j(t,r,z)$ and $j_0(t,r,z)$ given in Eqs.~(\ref{j0final2},\ref{jfinal2},\ref{jzfinal2}) and calculated using these $W$ fields at time $t=t_c+20$ is shown 
in Figs.~\ref{f:jzmove},\ref{f:jmove},\ref{f:j0move}, respectively. The current is confined to a region that extends over a distance of $\approx \pm 40$ into the $x-z$ plane and a distance of $\approx \pm 20$ along the longitudinal direction. This region is quite comparable to the bubble overlap region at this time as shown in Fig.~\ref{f:overlap}. 
In Fig.~\ref{f:jcmove} we show the current $j^\phi(x)$ appearing in Eq.~(\ref{maxazb1}) for the azimuthal magnetic field.

\begin{figure}
\centerline{\epsfig{file=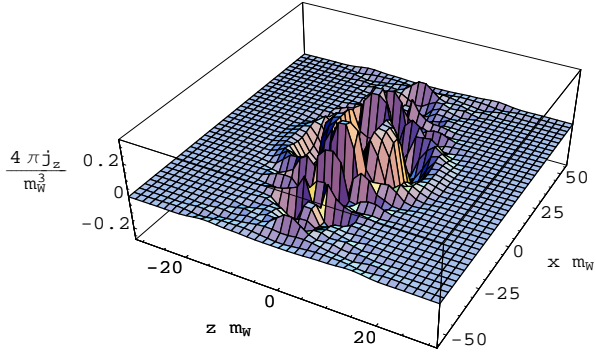,height=6cm,width=8.cm}}
\caption{ Current $4\pi j_z(t,\vec x)$ in the $x-z$ plane at $t= t_c+20$.}
\label{f:jzmove}
\end{figure}

\begin{figure}
\centerline{\epsfig{file=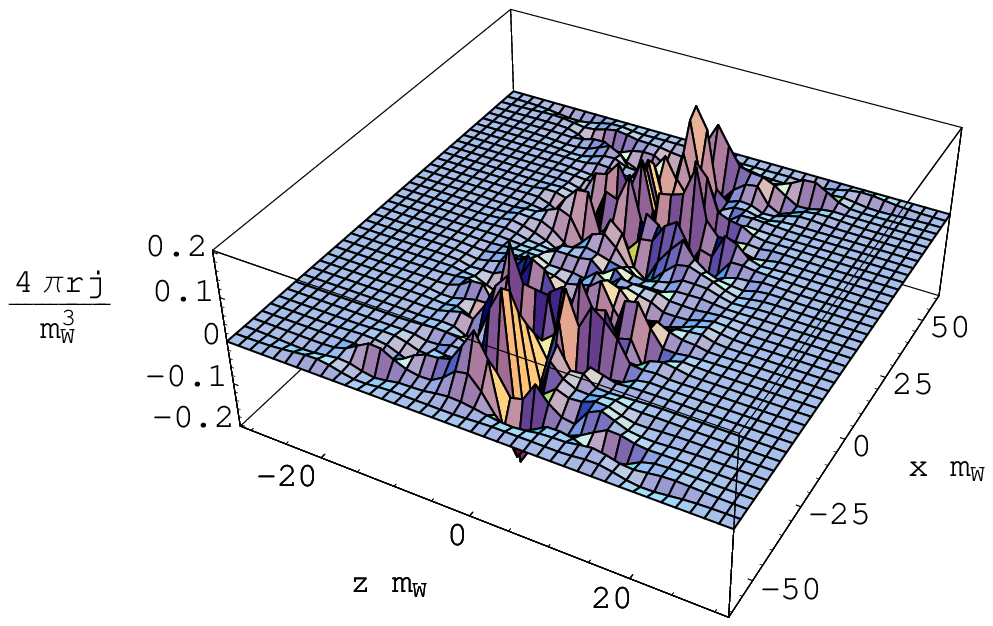,height=6cm,width=8.cm}}
\caption{ Current $4\pi rj(t,\vec x)$ in the $x-z$ plane at $t= t_c+20$.}
\label{f:jmove}
\end{figure}

\begin{figure}
\centerline{\epsfig{file=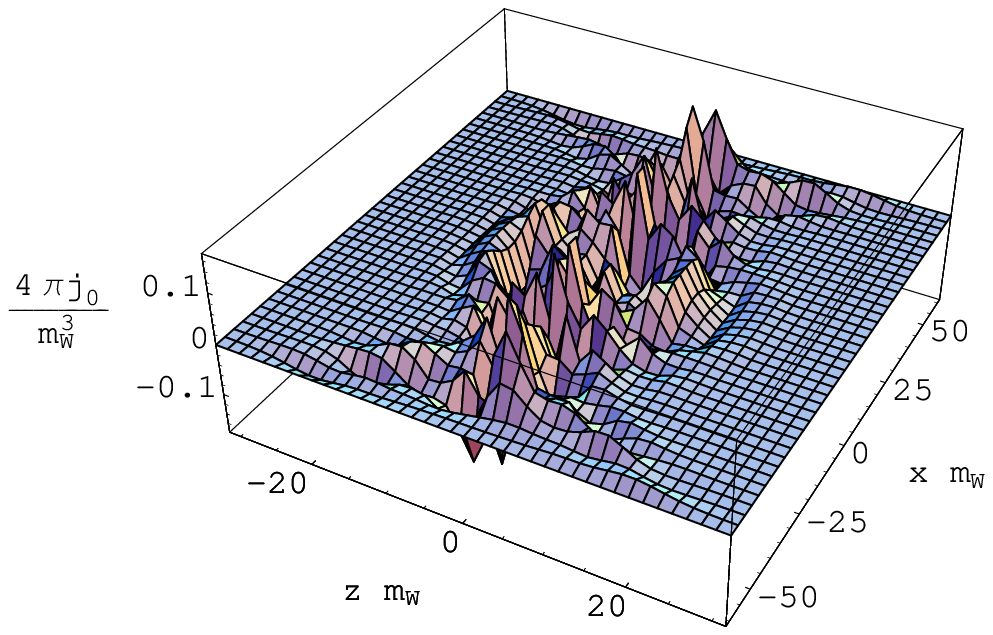,height=6cm,width=8.cm}}
\caption{ Current $4\pi j_0(t,\vec x)$ in the $x-z$ plane at $t= t_c+20$.}
\label{f:j0move}
\end{figure}

\begin{figure}
\centerline{\epsfig{file=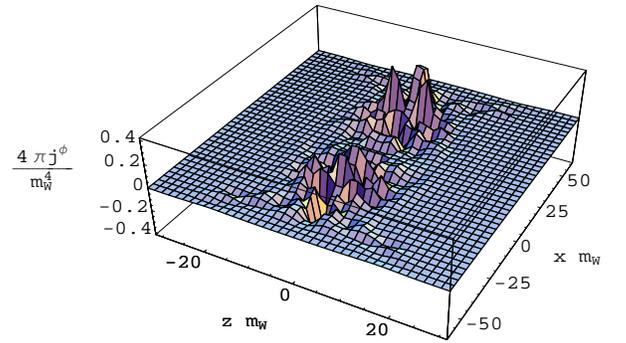,height=6cm,width=8.cm}}
\caption{ Current $4\pi j^\phi(t,\vec x)$ in the $x-z$ plane at $t= t_c+20$.}
\label{f:jcmove}
\end{figure}

\subsection{ Magnetic fields in bubble collisions, $a_s=4$ }
\label{s:thin}

The magnetic field calculated from Maxwell's Equation, Eq.~(\ref{maxazb1}), with the {\it em} current shown above and with boundary conditions as discussed is shown in Fig.~\ref{magpic3}. To facilitate comparison with the analogous results of Ref.~\cite{stevens1} with $O(1,2)$ symmetry it has been plotted at intervals $\delta t$ following the onset of the collision comparable to the results given there. As expected, the magnetic field moves away from $r=0$ and increases in magnitude with $t$. 
\begin{figure}
\centerline{\epsfig{file=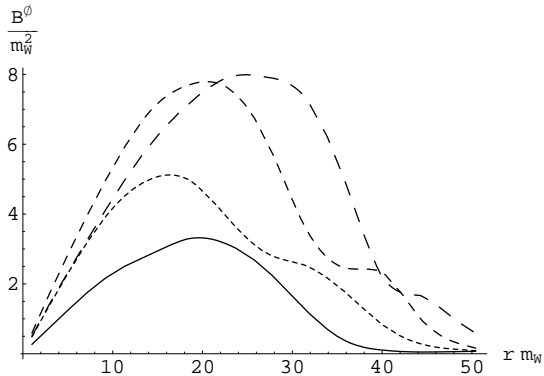,height=5cm,width=8.cm}}
\caption{ Magnitude of the azimuthal magnetic field in the transverse plane at $z$=0 as a function of $r$ for a series of times $t=t_c+\delta t$, where $\delta t=$5 (solid curve), 10 (short dash curve), 15 (medium dash curve) and, 20 (long dash curve). Field is calculated as in this work for $a_s=4$, $n_W'=1$, $v_{wall}=1$, and $\sigma =0$. }
\label{magpic3}
\end{figure}

The corresponding fields calculated from Ref.~\cite{stevens1} are shown in Fig.~\ref{oldmagpic2}. 
Comparing Figs.~\ref{magpic3},\ref{oldmagpic2} our magnetic field is about twice as large.  As in the $O(1,2)$ model, it is confined predominately to the region of bubble overlap shown in Fig.~\ref{f:overlap}.  The magnetic field is also apparently smoothed out by the surface so that our result in Fig.~\ref{magpic3} does not show the oscillations apparent in Fig.~\ref{oldmagpic2} and fills the region of bubble overlap more uniformly.

Thus, bubble surface dynamics seems to produce fields significantly larger in scale as well as magnitude. It is possible that bubble walls of even smaller surface thickness might grow even larger. We quantify this in Sect.~\ref{s:sens}. 
Calculations for the magnetic field including conductivity and wall speed $v_{wall}<1$ are given later in Sect.~\ref{s:sensvsig}.

\begin{figure}[tbh]
\centerline{\epsfig{file=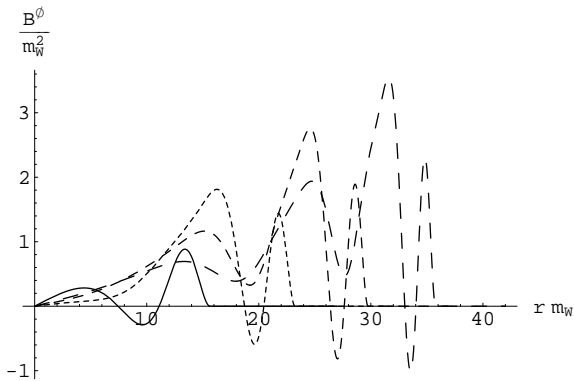,height=5cm,width=8.cm}}
\caption{Magnetic field calculated in Ref.~\cite{stevens1} in the transverse plane at $z$=0 as a function of $r$ for times  $t=t_c+\delta t$. Legend is the same as in Fig.~\ref{magpic3}. }
\label{oldmagpic2}
 \end{figure}

\begin{figure}
\centerline{\epsfig{file=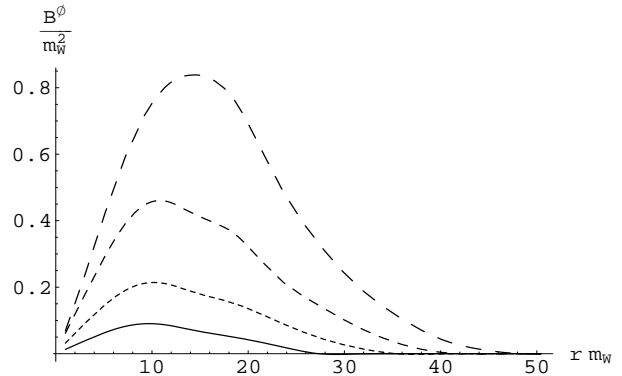,height=5cm,width=8.5cm}}
\caption{ Magnitude of the azimuthal magnetic field in the transverse plane at $z$=0 as a function of $r$.  Legend is the same as in Fig.~\ref{magpic3}. Field is calculated as in this work for $a_s=4$ and $n_W'=0$, $v_{wall}=1$, and $\sigma =0$. }
\label{magpic4}
\end{figure}

We close this section by illustrating the importance of the boundary condition  $\partial w_z(t_0)/\partial t$. Results for our present theory with $n_W'=0$ are shown in Fig.~\ref{magpic4}. Comparing Figs.~\ref{magpic3} and \ref{magpic4} it is seen that the magnitude and spatial scale of the magnetic field both decrease by setting $n_W'=0$. This can be understood as follows. There are two points to be made.  

First, as noted in Sect.~\ref{sss:BCond}, in the absence of a mechanism for $W$ of the plasma to enter the bubble as it expands ($n_W=0$), the initial speed of expansion $v_W(t=t_0)$ of the $W$ field is determined by the zero point motion of the $W$ in the bubble and its mass in the surface. With our choice of parameters, the momentum arising from the zero point motion is
\beq
p_w\sim \frac{1}{R_{0s}}=1/24\approx 0.042~,
\eeq
and the value of its mass in the surface is $\bar m_W=g\rho_{av} /\sqrt{2} \approx .0457$.  This suggests that as the expansion begins, 
\beq
v_W(t=t_0) &\approx& \frac{p_W}{\sqrt{p_w^2+\bar m_W^2}} \nonumber \\
&\approx& 0.68~,
\eeq
which is both smaller than the speed of the bubble wall, $v_{wall}=c$, and close to the speed of expansion the magnetic field seen in Fig.~\ref{magpic4}.  As time proceeds the spatial scale of the $W$ field, and hence the {\it em} current, increasingly lags the growth of the bubble wall.  Eventually, $\bar m_W$ approaches $m_W$ and the $W$ becomes even more confined to the interior of the bubble. 
The second point is that by choosing $n_W=1$ so that the number of $W$ populating the bubble grows in proportion to the volume displaced by the bubble, the {\it em} current naturally grows as well and leads to the larger magnitude seen in Fig.~\ref{magpic3}.

This discussion gives the rationale for the speculation made in Ref.~\cite{stevens3a} that the scale and magnitude of the magnetic field might be sensitive both to the boundary condition for $\partial w_z/\partial t$ as well as to the steepness of the bubble surface.

\section{Sensitivity of magnetic fields to bubble wall thickness}
\label{s:sens}

Since our theory accounts for surface dynamics, we expect its consequences to be quite different from earlier studies that did not address this aspect of the collisions. This expectation is based on the observation~\cite{stevens3a} that when the surface is considered new terms appear in the EOM that manifest a strong sensitivity to the steepness of the scalar field in the bubble surface. Based on the observations given here, one might expect the importance of the surface to grow as the surface becomes steeper and, conversely, less striking for a more diffuse bubble surface.

A quantitative measure of the sensitivity to the steepness of the scalar field in the surface may be obtained by comparing the calculation shown in Fig.~\ref{magpic3} to calculations with sharper walls. This sensitivity has, to our knowledge, never been studied quantitatively in previous work.
To obtain this quantitative measure, we compare here calculations with different surface diffuseness, $a_s=4$ and $a_s=3$. We show how the magnetic seed fields behave for $a_s=2$ by comparing to the results of Ref.~\cite{stevens3a}.

\begin{figure}[tbh]
\centerline{\epsfig{file=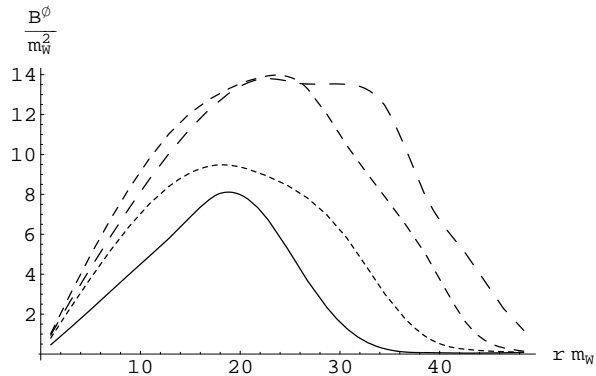,height=5cm,width=8.5cm}}
\caption{ Magnitude of the azimuthal magnetic field in the transverse plane at $z$=0 as a function of $r$ for a series of times  $t=t_c+\delta t$. Legend is the same as in Fig.~\ref{magpic3}.   Field is calculated as in this work for $a_s=3$ and $n_W'=1$, $v_{wall}=1$, and $\sigma =0$. }
\label{magpic2}
 \end{figure}

The calculation in Fig.~\ref{magpic2} is identical to the one in Fig.~\ref{magpic3}  but with $a_s=3$.  Comparing the magnetic fields in these two figures confirms that the peak size of the seed field $B_{Max}^\phi(t=t_c+20,a_s)/m_W^2$ increases as the wall becomes sharper.
In particular, decreasing $a_s$ from $a_s=4$ to $a_s=3$ results in nearly doubling this peak field. 

What about the peak field for $a_s=2$?  Comparing Fig.~\ref{magpic4} ($a_s=4$) to Fig.~3 of Ref.~\cite{stevens3a} ($a_s=2$), both of which are calculated with $n_W=0$, we see that decreasing $a_s$ from $a_s=4$ to $a_s=2$ results in nearly a 5-fold increase in the peak field.  
Since $B_{Max}^\phi(t=t_c+20,a_s=4)/m_W^2=8$ from Fig~\ref{magpic3}, we conclude that with $n_W=1$ the peak field would be $B^\phi_{Max}(t=t_c+20,a_s=2)/m_W^2=40$!

\section{Sensitivity of magnetic fields to $v_{wall}$ and conductivity}
\label{s:sensvsig}

Taking $v_{wall}<1$ and $\sigma \neq 0$ would of course modify the results found in the previous section. To quantify their effects in our model, we examine the case $a_s=4$, taking $v_{wall}=1/2$~\cite{hjk1}, and the same boundary conditions, shown in Fig.~\ref{f:bcIwz},\ref{f:bcIIwz}. 

We first show the sensitivity to $v_{wall}$ taking $\sigma=0$.  Results are shown in Fig.~\ref{bvwall}. To interpret them, it is useful to refer to Fig.~\ref{f:overlapv}, showing the region of bubble overlap at $t=t_c+20$ for $v_{wall}=1/2$. There are several points to make.

\begin{figure}
\centerline{\epsfig{file=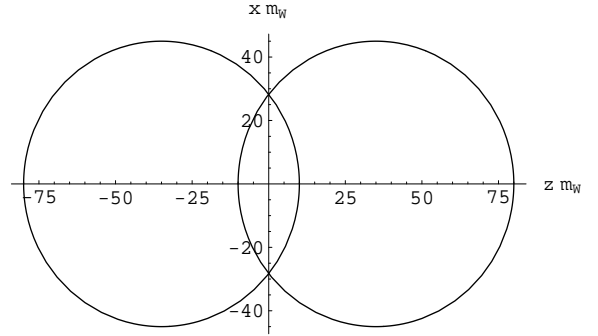,height=4.5cm,width=7.72cm}}
\caption{Showing the projection of the surface $S_c$ of the collision region onto the $x-z$ plane at time $t_f= t_c+20$ with $v_{wall}=1/2$. The surfaces $S_c$ is defined by Eq.~(\ref{boundsurf}) with $R_{1/2}(t_f)=55$.}
\label{f:overlapv}
\end{figure}

\begin{figure}
\centerline{\epsfig{file=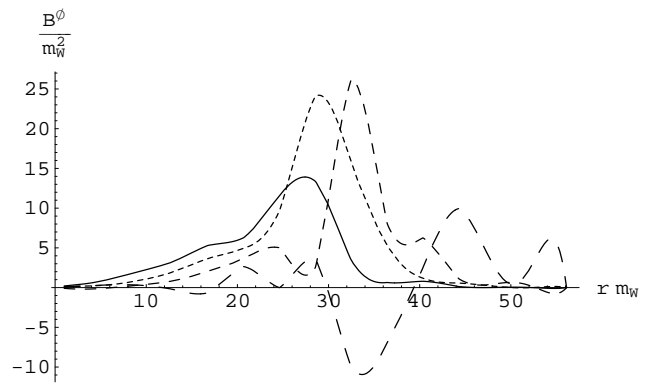,height=5cm,width=8.5cm}}
\caption{ Magnitude of the azimuthal magnetic field in the transverse plane at $z$=0 as a function of $r$ for a series of times  $t=t_c+\delta t$. Legend is the same as in Fig.~\ref{magpic3}.  Field is calculated as in this work for $a_s=4$ and $n_W'=1$, $v_{wall}=1/2$, and $\sigma =0$. }
\label{bvwall}
\end{figure}

First, note that the scale of the magnetic field in Fig.~\ref{bvwall} exceeds that in Fig.~\ref{magpic3}.  This is because the magnetic field propagates with the speed of light and thus quickly escapes the bubble into the surrounding plasma.  
Secondly, we see that the peak magnetic field is larger than the peak magnetic field in Fig.~\ref{magpic3}.  The reason for this behavior is that the current is due to the $W$ fields, which propagate in the plasma with an effective mass that is less than $m_W$; hence, even though the bubble surface propagates with a speed less than the speed of light, the current outside the bubble actually has a greater speed, allowing a larger magnetic field there.

\begin{figure}
\centerline{\epsfig{file=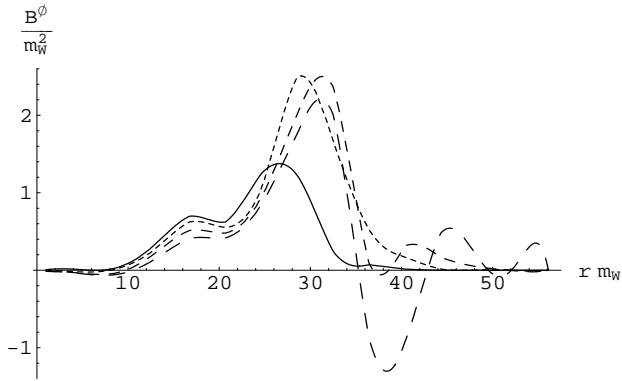,height=5cm,width=8.5cm}}
\caption{ Magnitude of the azimuthal magnetic field in the transverse plane at $z$=0 as a function of $r$ for a series of times  $t=t_c+\delta t$. Legend is the same as in Fig.~\ref{magpic3}.   Field is calculated as in this work for $a_s=4$ and $n_W'=1$, $v_{wall}=1/2$, and $\sigma =6.7T$. }
\label{bvsigwall}
\end{figure}

Next we show in Fig.~\ref{bvsigwall} the same calculation but with $\sigma =6.7T$ as used in Ref~\cite{ae98}, taking $T=T_c=199$GeV~\cite{eikr}.  It is again useful to refer to Fig.~\ref{magpic3}, as well as Fig.~\ref{bvwall}.  There are several points to note here as well.  First, the scale of the magnetic field has decreased in comparison to that in Fig.~\ref{bvwall}.  This has occurred at the expense of the field outside the bubble, a consequence of the dissipation characteristic $\sigma \neq 0$.  Thus, the field tends to be confined or ``frozen" in the bubble.  Both of these points are both consistent with the discussion of Sect.~\ref{medef}.  

The net effect of $v_{wall}<1$ and $\sigma \neq 0$  is to produce fields somewhat smaller than and of comparable spatial scale to those shown in Fig.~\ref{magpic3}, which was calculated in the absence of these effects.  The fact that there is such a small net effect is perhaps unexpected based on the experience with the Abelian Higgs model as discussed in Sect.~\ref{medef}.  The larger size found here has been explained from the physics of the current arising from the charged $W$ as opposed to that arising from the phase of the Higgs field:  in the latter case, the current is strictly confined to the region within the bubbles, but in the former case the current is not so restricted.

\section{Summary and Discussion}
Using EOM derived from the MSSM, we have explored magnetic seed field
creation during a first order primordial EWPT, focusing on the
role that bubble surface dynamics plays in creating such fields by extending the results of Ref.~\cite{stevens3a}. In our theory the charged gauge bosons $W^\pm$ of the non-Abelian sector and fermions, through the conductivity current, are the sources of the
{\it em} current producing the seed fields.  We
develop a linearized version of the theory applicable for the
case of gentle collisions, where the Higgs field is largely
unperturbed by the collision process~\cite{stevens1}.

In the theory developed here, the $W^{\pm}$ fields contributing to the {\it em} current evolve from initial conditions applied before the bubbles collide. These $W$ fields initially fill
the bubble uniformly and expand at the same speed as
the scalar field of the bubble containing them, consistent with the physical picture~\cite{stevens1} adopted in earlier
studies.

Because of our particular focus on the dynamics of the bubble surface, the EOM must be solved numerically to obtain the $W^\pm$ fields for nucleation and collisions. 
The magnetic field obtained from these solutions were found to be larger in both magnitude and scale than the corresponding one embodying $O(1,2)$ symmetry and jump boundary conditions at the time of collision~\cite{jkhhs,stevens1}. By comparing results calculated with surfaces of different slopes we found the seed fields to be quite sensitive to the bubble surface, and we obtained a quantitative measure of the sensitivity to the steepness of the bubble surface. These results help understand why our results are larger in scale and magnitude than those ignoring surface dynamics. 

We have not attempted to determine the present day magnetic fields
that are seeded by our fields generated during the EWPT.  This is a
complicated problem of plasma physics that has been studied
extensively.  It is known that the characteristics of the primordial
magnetic field are vastly modified during cosmic evolution to the
present day.  To model this evolution in magnetohydronamics,
one first solves the equations for the period leading to the
formation of galaxies and galactic clusters considering all relevant
dissipative processes such as viscosity, and then for the evolution
of these structures including the  possibility that they provide a
large-scale dynamo. Such studies have led to quantitative predictions
for magnetic field energy and coherence length at the present epoch.
The most recent of these~\cite{jed} support the possibility that
galactic cluster magnetic fields may be entirely primordial in origin.

Our present results reinforce the hope that the EWPT is a promising source for production of seed fields in the
MSSM for large-scale galactic and extra-galactic fields observed
today.  

\vspace{3mm}

\Large{{\bf Acknowledgements}}\\
\normalsize
The authors gratefully acknowledge Los Alamos National Laboratory for
its partial support of
the PhD research of T. Stevens. MBJ thanks the Department of Energy
for partial support.  We acknowledge helpful discussions with E. Henley and L. Kisslinger. 
\vspace{2cm}

\appendix

\section{The Scalar Field of a Bubble}
\label{a:scalarpar}

In this appendix we describe our scalar field $\rho(x)$ for a 
single bubble. This field is is taken to have the form
\beq
\label{sforms}
\rho(t,r,z)&=& \rho_c f^s(t,r,z)~,
\eeq
where $f^s(t,r,z)$ gives its shape and time-dependence, $\rho_c=\rho(t,0,0)$ is its value at the center of a bubble, and $f^s(t,0,0)=1$ fixes its normalization. 

As discussed in Sect.~\ref{ss:firstew}, in the absence of a medium $ \rho(x)$ is an instanton
solution of Coleman's equation, Eq.~({\ref{eomflead2}), at nucleation $t=t_n$. 
The magnitude of this solution is constant at $\rho_c=\rho_0$ to a high degree of accuracy throughout a 
region near the center of this bubble and then drops to zero at the surface. The scalar potential $\rho_0$ is determined by $m_W$ from Eq.~(\ref{wmass}) as
\beq
\label{mav0}
m_W^2=\frac{g^2}{2}\rho_0^2~. 
\eeq 
Taking $m_W$ and $g$ from Eq.~(\ref{cc}), we find from Eq.~(\ref{mav0}),
\beq
\label{mav2a}
\rho_0\approx 176~{\rm GeV}\approx 2.19m_W.
\eeq

Within a medium, many-body corrections to the scalar field may be taken into account by introducing an effective scalar field. In the absence of a detailed understanding, $f^s(t,r,z)$ is taken to have the surface characteristics of the Coleman solution in the so-called thin-wall approximation, 
\beq
\label{formsH}
f^s(t,r,z)= \frac{1}{f_s(t)} (1-\tanh\frac{R(r,z)- R_{0s}(t)}{a_s})~,
\eeq
where
\beq
\label{formsHn}
f_s(t)= 1-\tanh\frac{- R_{0s}(t)}{a_s}
\eeq
maintains the normalization and 
$\rho_c\to \rho_0-\rho_{av}$, where  $\rho_{av}$ accounts for the presence of the other bubbles on the average as discussed in Sect.~\ref{ss:medium}
Equation~(\ref{formsH}) has the structure of the scalar field in Refs.~\cite{stevens3a} and resembles the scalar field of Eq.~(6) in Sect. V of Ref.~\cite{cst00}. The functional form given in Eq.~(\ref{formsH}) is used to characterize the distribution of other constituents of the bubble as well.

In Eq.~(\ref{formsH}), $R(r,z)$ is the distance from the center of the bubble to
any point $\vec x =(r,z)$ in the medium, with $a_s$ determining the fall off of the scalar field in the bubble surface.  The function $R_{0s}(t)$ specifies the time-dependence of the bubble radius, 
\beq
\label{wallradius1}
R_{0s}(t)= R_t+v_{wall}(t-t_n)~,
\eeq
and is parametrized in terms of the asymptotic wall speed $v_{wall}$ and nucleation time $t_n$ of the bubble.

Defining $R_{1/2}(t)$ as the radius at which the scalar field is half its central value, we find from Eq.~(\ref{formsH},\ref{formsHn}) that
\beq 
\label{wallradius}
R_{1/2}(t) = \frac{a_s}{2}\log(2+\exp\frac{2 R_{0s}(t)}{a_s})~.
\eeq
The radius of the bubble at nucleation, $r_{ns}\equiv R_{1/2}(t_n)$, is then
\beq
r_{ns}= \frac{a_s}{2}\log(2+\exp\frac{2 R_t}{a_s})~,
\eeq
showing that $R_t\approx r_{ns}$ for large $R_t/a_s$.

With the nucleation centers for the bubbles located symmetrically on the z-axis at $z=\pm z_{0}$, the collision time $t_c$ for bubbles nucleated simultaneously may be found by solving the equation $R_{1/2}(t_c)=z_0$.  Taking $R_{0s}(t)$ from Eq.~(\ref{wallradius1}), 
\beq 
\label{tmfirst}
t_c= t_n+\frac{a_s}{2v_{wall}} \log[\frac{-2+e^{\frac{2z_0}{a_s}}}
{-2+e^{\frac{2r_{ns}}{a_s}}}]~.
\eeq

\section{Boundary condition on derivative of $w_z$}
\label{a:bcderwz}

Requiring that the rate at which  $W^{\pm}$ quanta enter the bubble be the same as the rate at which they leave the modes from which they originated in the thermal plasma, we have
\beq
\label{nprimec}
\frac{d}{dt} N^W_{bubble}(t) &=& \frac{d}{dt} N^W_{plasma}(t)~.
\eeq
This condition maintains the average density of $W$ inside the bubble roughly constant as a function of time. 

The quantity $N^W_{plasma}$ is given by $N^W_{plasma}=\rho_W(T)V_{bubble}$, where $V_{bubble}$ is the volume of the bubble and $\rho_W(T)$ is the density of the $W^{\pm}$ in the plasma at temperature $T$ that are able to make the transition into the bubble. Thus, the right-hand side of Eq.~(\ref{nprimec}) may be written
\beq
\label{nprimeca}
\frac{d}{dt} N^W_{plasma}(t) 
&=&\rho_W(T) \frac{d}{dt} V_{bubble}~,
\eeq
where $d/dtV_{bubble}$ is the rate at which the bubble displaces the plasma volume.

Since the volume of the bubble is the same as the volume occupied by its scalar field, we may write
\beq
\label{nprimec0}
 V_{bubble}(t) &=& \int f^s(t,\vec x)^2 dV~.
\eeq
Then, taking $\rho^W(T) =2m_Wn_W^2$ from Eq.~(72) of the Appendix of Ref.~\cite{stevens1}, with $\sqrt{2m_W}$ the relativistic normalization of the $W$ wave function, we find \footnote {To avoid confusing the notation for the magnitude of $w_z(t_0,\vec x)$ with other components of the $W$ field, we call this magnitude here $n_W$ (called $w$ in Ref.~\cite{stevens1}).}
\beq
\label{nprimec1}
\rho^W(T)\frac{d}{dt} V_{bubble} &=& 2m_W n_W^2 \int \frac{\partial f^s(t,\vec x)^2}{\partial t} dV~.
\eeq

Likewise, $N^W_{bubble}$ is determined as 
\beq
\label{nprimec2b}
N^W_{bubble}(t) &=& 2m_W\int w_z(t,\vec x)^2 dV~.
\eeq
Thus, the left-hand side of Eq.~(\ref{nprimec}) may written 
\beq
\label{nprimec2a}
\frac{d}{dt} N^W_{bubble}(t)&=& 4m_W \nonumber \\
&\times&\int w_z(t,\vec x) \frac{\partial w_z(t,\vec x)}{\partial t}dV ~.
\eeq
or
\beq
\label{nprimec2}
\frac{d}{dt} N^W_{bubble}(t)&=& 4  n_W n_W'm_W \nonumber \\
&\times&\int f^w(t,\vec x) \frac{\partial f^w(t,\vec x)}{\partial t}dV ~.
\eeq
To obtain Eq.~(\ref{nprimec2}) have used Eq.~(\ref{wLRdef}) and taken the boundary condition on the time derivative of the $w_z$ field at $t\approx t_0$ to be determined by the same function that determines the shape of $w_z(t_0,\vec x )$ in Eq.~(\ref{wLRdef}), given in Eq.~(\ref{wzLRderdef}).

The normalization $n_W'$ in Eq.~(\ref{nprimec2}) may now be determined by equating Eqs.~(\ref{nprimec1},\ref{nprimec2}) at $t=t_0$, with the result 
\beq
\label{nwprime}
n_W'=n_W \frac{\int \partial f^{s2}(t_0,\vec x)/\partial t dV}{ \int \partial f^{w2}(t_0,\vec x) /\partial t dV}~.
\eeq

\section{Constrained initial conditions in cylindrical coordinates}
\label{a:initcondA}

In this Appendix we express the constrained initial 
conditions discussed in Sect.~\ref{ss:initcond} in cylindrical coordinates using the specific form of the $W$ field at $t=t_0$ motivated and discussed in Sect.~\ref{sss:BCond}.

Referring to 
Eqs.~(\ref{kge},\ref{bc1},\ref{bc2}) the auxiliary condition in cylindrical coordinates is enforced by expressing $\chi^a(x)$, defined in 
Eq.~(\ref{chidef}), and 
$\partial \chi^a( x)/\partial t $, with the aid of Eqs.~(\ref{cyn1},\ref{cyn2}). This gives
\beq
\label{chi}
\chi^a(x) &=& \frac{\partial w^{a}_0 }{\partial t}
+2w^a+r\frac{\partial w^a }{\partial r} - \frac{\partial w^a_z
}{\partial z} \nonumber \\
&+& 2(w_0^a \frac{\partial \ln m}{\partial t}
+rw^a \frac{\partial \ln m}{\partial r} \nonumber \\
&-&w^a_z\frac{\partial \ln m}{\partial z})
\eeq
and
\beq
\label{chider}
\frac{\partial \chi^a(x)}{\partial t} &=& \frac{\partial}{\partial t}
(\frac{\partial w^{a}_0 }{\partial t} +2w^a+r\frac{\partial w^a
}{\partial r} - \frac{\partial w^a_z }{\partial z}) \nonumber \\
&+&2\frac{\partial}{\partial t}( ( w_0^a \frac{\partial \ln m}{\partial t}
+rw^a \frac{\partial \ln m}{\partial r} \nonumber \\
&-&w^a_z\frac{\partial \ln m}{\partial z})
)~.
\eeq
Equation~(\ref{chider}) may be equivalently expressed as
\beq
\label{chider5}
\frac{\partial \chi^a(x)}{\partial t} &=& \frac{\partial}{\partial t}
(2w^a+r\frac{\partial w^a }{\partial r} - \frac{\partial w^a_z
}{\partial z}) +\frac{\partial^2 w^{a}_0}{\partial
r^2}\nonumber \\
&+&\frac{1}{r}\frac{\partial w^{a}_0}{\partial r}
+\frac{\partial^2 w^{a}_0}{\partial z^2} -m^2 w^{a}_0~,
\eeq
by using the EOM, Eq.~(\ref{eqw1a}),
and canceling the term $\partial^2 w^{a}_0/\partial t^2 $.

Using Eqs.~(\ref{chi},\ref{chider5}), the conditions that
$\chi^a(t=t_0,\vec x)=0$ and $\partial \chi^a(t=t_0, \vec x)/\partial t=0
$ become, respectively,
\beq
\label{chi0}
0  &=& (\frac{\partial w^{a}_0 }{\partial t}
+2w^a+r\frac{\partial w^a }{\partial r} - \frac{\partial w^a_z
}{\partial z}) \nonumber \\
&+& 2( w_0^a \frac{\partial \ln m}{\partial t}
+rw^a \frac{\partial \ln m}{\partial r} \nonumber \\
&-&w^a_z\frac{\partial \ln m}{\partial z})
\eeq
and
\beq
\label{chider0}
0 &=& 2\frac{\partial w^a}{\partial t} +r\frac{\partial^2 w^a
}{\partial r\partial t} - \frac{\partial^2 w^a_z }{\partial z\partial
t} \nonumber \\
&+&\frac{\partial^2 w^{a}_0}{\partial
r^2}+\frac{1}{r}\frac{\partial w^{a}_0}{\partial r}
+\frac{\partial^2 w^{a}_0}{\partial z^2} -m^2 w^{a}_0~.
\eeq
In accord with Sect.~\ref{ss:ourmodel}, at $t=t_0$ for both nucleation and collisions, the $W$ quanta occupy 
$w^a_z$ with the other components $w_0^a$ and $w^a$ empty.  Imposing the boundary conditions $w^a(t_0,\vec x)=0$ and $w^a_0(t_0,\vec x)=0$,   
Eqs.~(\ref{chi0},\ref{chider0}) simplify and become
\beq
\label{chi3}
0 &=& \frac{\partial w^{a}_0(t_0,\vec x) }{\partial
t} - \frac{\partial w^a_z(t_0,\vec x)
}{\partial z} \nonumber \\
&-&2w^a_z(t_0,\vec x)\frac{\partial \ln m(t_0,\vec x)}{\partial z}
\eeq
and
\beq
\label{chider3}
0 &=& 2\frac{\partial w^a(t_0,\vec x)}{\partial t} +r\frac{\partial^2
w^a(t_0,\vec x) }{\partial r\partial t} \nonumber \\
   &-& \frac{\partial^2 w^a_z(t_0,\vec x) }{\partial z\partial t}~.
\eeq
With $m(t,\vec x)$ a continuous function, Eqs.~(\ref{chider3},\ref{chi3}) as well as the auxiliary condition Eq.~(\ref{aux1}) are satisfied.

\subsection{ Collisions }
\label{aa:collisions}

The colliding bubbles, which are nucleated simultaneously at time $t=t_n$ at points $z=\pm z_0$ located symmetrically about the origin on the z-axis with nucleation radii $r_{ns}$, are naturally assumed to be well separated and non-overlapping before they collide, as required by the discussion in Sect.~\ref{ss:colkin}.  Thus, the boundary conditions and initial conditions on the $W$ fields are established well before the collision, satisfying the requirements of Sect.~\ref{ss:dyniss}.

The functions $w_{L}$, $w_{R}$ and their time derivatives depend on $r$ and $z$ entirely through $R=\sqrt{r^2+(z\pm z_0)^2}$ at $t=t_0$, just as for the scalar field (see Eqs.~(\ref{wLRdef},\ref{wzLRderdef})). Thus,
\beq
\label{wzder1}
\frac{\partial w_{z\pm}(r,z)}{\partial z}&=&\frac{z\pm z_0}{R}\frac{\partial}{\partial R} w_{z\pm} \nonumber \\
&=&\frac{z\pm z_0}{r}\frac{\partial}{\partial r} w_{z\pm } ~,
\eeq
and
\beq
\label{wzder2}
\frac{\partial }{\partial z}\frac{w_{z\pm}(r,z)}{\partial t}&=&\frac{z\pm z_0}{R}\frac{\partial}{\partial R} \frac{\partial w_{z\pm} }{\partial t} \nonumber \\
&=&\frac{z\pm z_0}{r}\frac{\partial}{\partial r} \frac{\partial w_{z\pm }}{\partial t} ~.
\eeq
where the positive sign corresponds to the left-hand bubble and the negative sign the right-hand bubble.

The initial condition on $\partial w^a/\partial t$ is found from
the auxiliary condition in Eq.~(\ref{chider3}),
\beq
\label{chider2}
(2 +r\frac{\partial }{\partial r})\frac{\partial w^a }{\partial t}
&=& \frac{1}{r} \frac{\partial }{\partial r}r^2\frac{\partial w^a }{\partial t} \nonumber \\
&=&\frac{\partial}{\partial z} \frac{\partial w^a_z }{\partial t}
~,
\eeq
which determines $\partial
w^a(t=t_0,\vec x)/\partial t$ in terms of $\partial w^a_z(t=t_0,
\vec x) /\partial t$.
Using Eq.~(\ref{wzder2}) and the boundary conditions on
$w^a_z(t_0,\vec x)$ in Eqs.~(\ref{bcI},\ref{bcII}), Eq.~(\ref{chider2}) becomes
\beq
\label{chider2a}
\frac{1}{r} \frac{\partial }{\partial r}r^2\frac{\partial w^a }{\partial t} 
&=&\frac{z\pm z_0}{r}\frac{\partial}{\partial r} \frac{\partial w^a_{z\pm }}{\partial t}
~.
\eeq
This may be integrated immediately to give 
\beq
\label{specII0}
\frac{\partial }{\partial t}  w^{a}(t_0, \vec x) &=& \frac{
z\pm z_0 } {r^2}( \frac{\partial }{\partial t} w^a_{z\pm }(r,z) \nonumber
\\
&-& \frac{\partial }{\partial t} w^a_{z\pm }(r=0,z) )  ~.
\eeq
For $a=$II we find
\beq
\label{specII}
\frac{\partial }{\partial t}  w^{II}(t_0, \vec x) &=& \frac{
z+z_0 } {r^2}( \frac{\partial }{\partial t} w_{L}(r,z) \nonumber
\\
&-& \frac{\partial }{\partial t} w_{L}(r=0,z) ) \nonumber \\
&+& \frac{ z-z_0 }{r^2}( \frac{\partial }{\partial t} w_{R}(r,z)
\nonumber \\
&-& \frac{\partial }{\partial t} w_{R}(r=0,z))
\eeq
and for $a=$I
\beq
\label{specI}
\frac{\partial }{\partial t}  w^{I}(t_0, \vec x) &=&   \frac{
z+z_0}{r^2}( \frac{\partial }{\partial t} w_{L}(r,z) \nonumber \\
&-& \frac{\partial }{\partial t} w_{L}(r=0,z) )\nonumber \\
&-& \frac{ z-z_0 }{r^2}( \frac{\partial }{\partial t} w_{R}(r,z)
\nonumber \\
&-& \frac{\partial }{\partial t} w_{R}(r=0,z))~.
\eeq

Next, using Eq.~(\ref{chi3}) and once again the boundary conditions on
$w_z(t_0,\vec x)$ in Eqs.~(\ref{bcI},\ref{bcII}), we find ${\partial w_0}/{\partial t}$ for
BCII to be
\beq
\label{bcII0t}
\frac{\partial w_0^{II}(t_0, \vec x)}{\partial t} &=&\frac{\partial
}{\partial z} w^{II}_z(t_0, \vec x)  +2w_z^{II}(t_0,\vec x)\nonumber \\
&&\times \frac{\partial}{\partial z}\ln m(t_0, \vec x)
\eeq
and for BCI to be
\beq
\label{bcI0t}
\frac{\partial w_0^{I}(t_0, \vec x)}{\partial t}&=&\frac{\partial
}{\partial z} w^{I}_z(t_0, \vec x) +2w_z^{I}(t_0,\vec x)\nonumber \\
&&\times \frac{\partial}{\partial z}\ln m(t_0, \vec x)
~.
\eeq

\subsection{Nucleation}
\label{aa:nucleation}

Since the positively and negatively charged 
$W$ field evolve exactly the same in a single bubble with our initial conditions, the distinction between the charged fields is immaterial for 
nucleation and we drop the superscript on $w^a(t,r,z)$ 
that distinguishes between $W^+$ and $W^-$. 
Then, assuming that the bubble is nucleated at the origin, $r=z=0$, with nucleation radius $r_{ns}$, and assuming that the scalar field of the bubble expands in accord with Eq.~(\ref{wallradius1}), the initial conditions for $t_n \approx t_0<< t_c$ are found as follows.

For $\partial w/\partial t$, the initial condition is found from
Eq.~(\ref{chider3}) with the boundary conditions on
$w_z(t_0,\vec x)$ in Eqs.~(\ref{bcIc11n}).  As in Eqs.~(\ref{specII},\ref{specI}),  we find
\beq
\label{spec0n}
\frac{\partial }{\partial t}  w(t_0, \vec x) &=& \frac{
z } {r^2}( \frac{\partial }{\partial t} w_{z }(r,z) \nonumber
\\
&-& \frac{\partial }{\partial t} w_{z }(r=0,z) )  ~. 
\eeq
Finally, using Eq.~(\ref{chi3}) and the boundary conditions on
$w_z(t_0,\vec x)$ in Eq.~(\ref{bcIc11n}), we find ${\partial w_0}/{\partial t}$ as in Eqs.~(\ref{bcII0t},\ref{bcI0t}),
\beq
\label{nuc0t}
\frac{\partial w_0(t_0, \vec x)}{\partial t} &=&\frac{\partial
}{\partial z} w_z(t_0, \vec x)  +2w_z(t_0,\vec x)\nonumber \\
&&\times \frac{\partial}{\partial z}\ln m(t_0, \vec x)~.
\eeq

\end{document}